\documentclass[fleqn,usenatbib]{mnras}

\usepackage{newtxtext,newtxmath}
\usepackage[T1]{fontenc}

\usepackage{graphicx}	
\usepackage{amsmath}	
\usepackage{siunitx}

\usepackage[utf8]{inputenc}\DeclareUnicodeCharacter{2212}{-}

\usepackage{etoolbox}

\makeatletter
\patchcmd\@combinedblfloats{\box\@outputbox}{\unvbox\@outputbox}{}{%
}%
\makeatother

\newcommand{\ie}[0]{\textit{i.e.} }
\newcommand{\eg}[0]{\textit{e.g.} }

\newcommand{\ra}[0]{\mathrm{R.A.}}
\newcommand{\dec}[0]{\mathrm{Dec.}}

\newcommand{\rc}[1]{#1}

\title[Angular clustering properties of the DESI QSO TS]{Angular clustering properties of the DESI QSO target selection using DR9 Legacy Imaging Surveys}

\author[E. Chaussidon et al.]{
Edmond Chaussidon$^{1}$\thanks{E-mail: edmond.chaussidon@cea.fr},
Christophe Y\`eche$^{1}$,
Nathalie Palanque-Delabrouille$^{1}$,
\newauthor
Arnaud de Mattia$^{1}$,  Adam D.\ Myers$^{2}$,  Mehdi Rezaie$^{3}$,  Ashley J. Ross$^{4}$,  Hee-Jong Seo$^{3}$, 
\newauthor
David Brooks$^{5}$,  Enrique Gazta\~naga$^{6}$,  Robert Kehoe$^{7}$,  Michael E. Levi$^{8}$,  
\newauthor
Jeffrey A. Newman$^{9,10}$,  Gregory Tarl\'e$^{11}$,  Kai Zhang$^{8}$\\
\\
$^{1}$IRFU,  CEA,  Universit\'e Paris-Saclay,  F-91191 Gif-sur-Yvette,  France\\
$^{2}$Department of Physics \& Astronomy,  University of Wyoming, 1000 E. University Ave., Laramie, WY 82071, USA\\
$^{3}$Department of Physics and Astronomy, Ohio University, Athens, OH 45701,  USA\\
$^{4}$Center of Cosmology and AstroParticle Physics,  The Ohio State University,  Columbus,  OH 43210,  USA\\
$^{5}$Department of Physics \& Astronomy,  University College London,  Gower Street,  London,  WC1E 6BT,  UK\\
$^{6}$Institute of Space Sciences (ICE, CSIC),  Campus UAB,  Carrer de Can Magrans,  sn,  08193 Bellaterra (Barcelona),  Spain\\
$^{7}$Department of Physics,  Southern Methodist University,  3215 Daniel Ave., Dallas,  TX,  75205,  USA\\
$^{8}$Lawrence Berkeley National Laboratory,  1 Cyclotron Road, Berkeley,  CA 94720,  USA\\
$^{9}$Department of Physics and Astronomy,  University of Pittsburgh,  Pittsburgh,  PA 15260,  USA\\
$^{10}$Pittsburgh Particle Physics,  Astrophysics,  and Cosmology Center (PITT PACC),  University of Pittsburgh,  Pittsburgh,  PA 15260, USA\\
$^{11}$Department of Physics,  University of Michigan,  Ann Arbor,  MI 48109, USA
}
\date{Accepted XXX. Received YYY; in original form ZZZ}

\pubyear{2021}

\begin{document}
\label{firstpage}
\pagerange{\pageref{firstpage}--\pageref{lastpage}}
\maketitle

\begin{abstract}
The quasar target selection for the upcoming survey of the Dark Energy Spectroscopic Instrument (DESI) will be fixed for the next five years. The aim of this work is to validate the quasar selection by studying the impact of imaging systematics as well as stellar and galactic contaminants, and to develop a procedure to mitigate them. Density fluctuations of quasar targets are found to be related to photometric properties such as seeing and depth of the Data Release 9 of the DESI Legacy Imaging Surveys. To model this complex relation, we explore machine learning algorithms (Random Forest and Multi-Layer Perceptron) as an alternative to the standard linear regression. Splitting the footprint of the Legacy Imaging Surveys into three regions according to photometric properties, we perform an independent analysis in each region, validating our method using eBOSS EZ-mocks. The mitigation procedure is tested by comparing the angular correlation of the corrected target selection on each photometric region to the angular correlation function obtained using quasars from the Sloan Digital Sky Survey (SDSS) Data Release 16. With our procedure, we recover a similar level of correlation between DESI quasar targets and SDSS quasars in two thirds of the total footprint and we show that the excess of correlation in the remaining area is due to a stellar contamination which should be removed with DESI spectroscopic data. We derive the Limber parameters in our three imaging regions and compare them to previous measurements from SDSS and the 2dF QSO Redshift Survey.
\end{abstract}

\begin{keywords}
survey - cosmology: observations – large-scale structures – dark energy
\end{keywords}


\section{Introduction}
The Dark Energy Spectroscopic Instrument\footnote{\url{https://www.desi.lbl.gov/}} (DESI) will conduct a spectroscopic survey of a third of  the sky (\ie $14{,}000~\rm{deg}^2$) over a period of 5 years \citep{DESICollaboration2016}.  At the end of the survey, DESI is expected to produce a catalogue of 35 million objects covering  redshifts up to $\sim 4$.  This will be the largest catalogue of spectroscopic sources ever observed, increasing by an order of magnitude the number of sources observed by the  Sloan Digital Sky Survey\footnote{\url{https://www.sdss.org/dr16/}} \citep[SDSS;][]{Ahumada2020}.

The DESI survey will provide the most detailed 3D map of the matter distribution in the Universe as traced by four different target classes: galaxies from a Bright Galaxy Survey (BGS) in the redshift range $0.05<z<0.4$,  Luminous Red Galaxies (LRGs) in the range $0.4<z<1.1$,  Emission Line Galaxies (ELGs) in the range $0.6<z<1.6$ and quasars or quasi-stellar objects (QSOs) at $z>0.8$. The DESI collaboration will use these datasets to perform analyses of baryon acoustic oscillations and redshift space distortions to constrain the expansion history of the Universe and the growth of structure.  Similar studies were conducted on the final SDSS dataset (Data Release 16; DR16) by \cite{Gil-Marin2020,  Wang2020,  Bautista2021} (using LRGs), \cite{Tamone2020,  Raichoor2021,  DeMattia2021} (using ELGs), \cite{Neveux2020,  Hou2021} (using QSOs) and \cite{DuMasdesBourboux2020} (using the Ly-$\alpha$ forest). DESI should also enable precise measurements of the sum of the neutrino masses, as well as investigations of modified gravity and theories of inflation. 

A promising approach to probe inflation is through the tiny imprint left on the matter power spectrum by inflation-induced primordial non-Gaussianity.  As noted for instance in \cite{Ross2013,  Castorina2019,  Mueller2021} this measurement is known to be limited by systematic effects on large scales,  most of which are due to imaging systematics  imprinted on the density of spectroscopic targets. To prepare for upcoming clustering analyses with DESI,  it is therefore crucial to study and be able to mitigate the imaging systematics that impact target selection. 

The aim of this paper is to mitigate systematics for the DESI QSO target selection. We compute the angular clustering properties of the QSO targets in order to validate the selection method and to provide a control sample of QSO targets stripped of residual biases from imaging systematics or selection criteria.  This is particularly important as the QSO target selection is known to be strongly contaminated by stars in addition to being impacted by imaging systematics.  This work also serves as a crucial input to the selection of the QSO targets for DESI,  which will soon be finalized for the five-year duration of the survey, \rc{ in order to avoid strong imprints into the spectroscopic QSO sample due to imaging systematics that occur during the target selection. } \rc{Finally,  it is also a proof of concept.  Since the stellar contamination of the target selection will be removed after the spectroscopic survey,  the same method could be used to mitigate systematics in the DESI spectroscopic QSO sample.} 

Different strategies have been developed to deal with  imaging systematic effects and improve the reliability of clustering studies.  In this work, we follow the approach that was used for SDSS studies \citep{Myers2006,  Ross2011,  Ho2012,  Ross2017,  Ross2020,  Raichoor2021} and \rc{for Dark Energy Survey \citep[][]{DESCollaboration2021} studies} \citep[][]{Leistedt2016, Elvin-Poole2018}.  This method models the variation of target density as a \rc{linear} function of imaging features \citep[see, \eg][]{Myers2015, Prakash2016} in order to remove fluctuations caused by imaging systematics.  A correction weight is then computed and applied to the data.  Since this method smooths density fluctuations, one needs to check to what level the mitigation procedure affects the cosmological signal. Another less common approach is based on mode projection \citep{Rybicki1992, Tegmark1998,  Leistedt2013,  Elsner2016,  Kalus2019}: Modes (in Fourier space) or pixels (in configuration space) are assigned increased variance where the systematics map exhibits large values,  such that the covariance matrix has larger values in the presence of systematics.  This is a robust method,  which,  however,  only mitigates systematics using \rc{also} a linear combination of the imaging maps.  It cannot model the non-linear effects that are now observed, as illustrated in \cite{Ho2012}.  The correction-weights and mode-projection strategies can be combined in a common framework as explained in \cite{Weaverdyck2021}.

To circumvent the assumption that \rc{only} known features can completely explain all imaging systematics, one can also use a forward-modeling approach \citep{Suchyta2016,  Burleigh2018,  Kong2020}. Such an approach accounts for source detection and target selection processes in an end-to-end fashion, by injecting fake galaxies into raw images,  running source detection on the images, and applying target selection algorithms to the resulting sources. 

Some studies of how imaging systematics affect DESI target selection have already been undertaken.  For instance, \cite{Kitanidis2020} gave a first overview of imaging systematics for different DESI target classes selected from Data Release 7 (DR7) of the DESI Legacy Imaging Surveys \citep[see e.g.][]{Dey2019},  and found that the DESI QSO target sample suffered from strong contaminating effects.  In another study of DR7 of the the Legacy Imaging Surveys,  \cite{Rezaie2019} analysed how imaging systematic affect the eBOSS-like ELG selection, and used an artifical neural network to mitigate non-linear effects.

The aim of this work is to analyse the final QSO target selection that will be used for the next five years by DESI.  The  selection is based on the ninth data release (DR9) of the Legacy Imaging Surveys, which is greatly improved compared to  prior data releases. For instance,  DR9 covers the full footprint of the Legacy Imaging Surveys (about 20{,}000 deg$^2$),  incorporates improved background-fitting,  and uses the latest definition of bright-star and large-galaxy masks.  To mitigate systematic effects on the QSO selection,  we will explore machine-learning approaches based on Random Forests (RFs) and Neural Networks (NNs),  which we will compare to each other as well as to a traditional linear treatment.   We will test our mitigation process by measuring the angular properties of the QSO target selection, before and after applying the weights derived from these techniques.  \rc{The density fluctuation of the QSO target selection is non-linear with respect to imaging properties due to significant stellar contamination. } We account for this contamination by including the Milky way and Sagittarius Stream stellar distributions as \rc{additional} inputs to our mitigation procedure. 

This paper is organised as follows.  In Section~\ref{sec:data},  we outline the DESI DR9 Legacy Imaging Surveys and introduce the strategy used QSO target selection. The observational features we study and the method used to mitigate systematics are detailed in Section~\ref{sec:method}.  In Section~\ref{sec:mitigation},  we show the results of our systematic mitigation and discuss how density fluctuations in the QSO target sample are related to observational features.  In Section~\ref{sec:clustering}, we analyse the angular clustering properties of the DESI QSO target sample,  illustrating the efficiency of our methods.  Finally,  our conclusions are presented in Section~\ref{sec:conclusion}.

All magnitudes in this paper will be quoted on the AB system, including magnitudes from the Wide-field Infrared Survey Explorer which are often given on the Vega system. In addition,  except when mentioned otherwise,  all computations with \texttt{HEALPix} pixels are done with $N_{\rm side}=256$ (a pixel area of $\sim$0.05 deg$^2$) and all maps are \rc{plotted} in a Mollweide projection with a \texttt{HEALPix} resolution of $N_{\rm side}=64$.

\section{Quasar targets in Legacy Imaging Surveys}
\label{sec:data}
In this section, we introduce the Legacy Imaging Surveys that are used to perform the DESI QSO target selection. We then present the QSO target selection method itself, and finally discuss areas of the DESI footprint that have particularly high target density.

\subsection{Data Release 9 of Legacy Imaging Surveys}
\label{sec:DR9}

\begin{table}
\centering
\caption{Median values of the PSF depth and PSF size (cf.  Section~\ref{sec:features}) for the three imaging surveys that together constitute the DR9 Legacy Imaging Surveys.  \rc{DECaLS is split according to the DES region since the quality of the photometry inside this region is significantly better. }}
\label{tab:quality_photometry}
\begin{tabular}{lcccccc}
\hline
                    				& \multicolumn{3}{c}{PSF Depth [mag]} & \multicolumn{3}{c}{PSF Size [arcsec]} \\
                				     & $g$   & $r$  & $z$  & $g$    &   $r$ & $z$       \\ \hline
DECaLS (non DES)    &     24.7     &     24.2     &    23.3     &    1.51      &   1.38      &   1.31      \\
DES              			    &      25.2    &     25.0     &    23.8     &     1.42     &   1.24      &    1.14     \\
BASS           			    &      24.2    &    23.7      &        &   1.89       &   1.67        &       \\
MzLS          		        &          &          &    23.3     &          &         &       1.24  \\ \hline
\end{tabular}
\end{table}

\begin{figure}
	\includegraphics[scale=1]{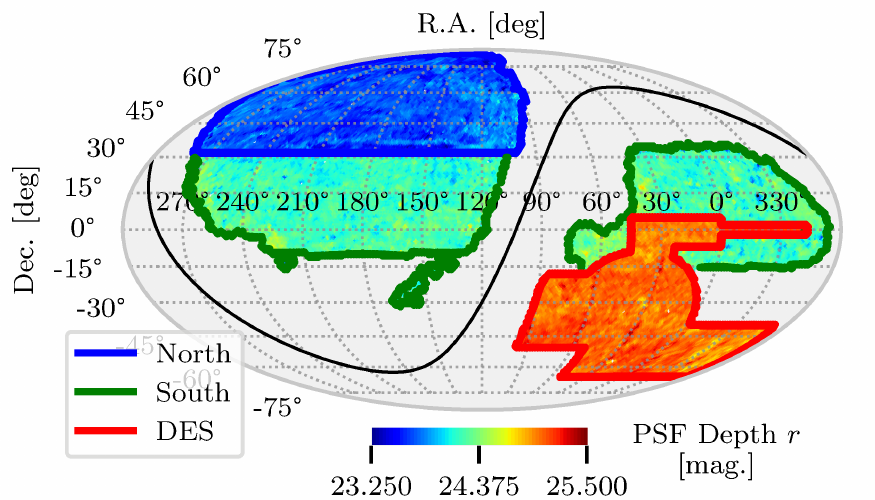}
    \caption{Distribution of the PSF Depth $r$ (cf.  Section~\ref{sec:features}) in the DR9 Legacy Imaging surveys footprint.  The $r$-band is used to define the magnitude limit for DESI QSO target selection. The solid black line shows the Galactic plane.  Three different imaging footprints are highlighted.  The blue region is the combination of BASS and MzLS
    (designated {\em North} in this paper). The red region is the DES part of DECaLS (designated {\em DES}).  The green region, which excludes the red and the blue regions, is the non-DES part of DECaLS (designated {\em South}).}
    \label{fig:psfdepth_r}
\end{figure}

To select targets for the DESI spectroscopic survey,  the Legacy Imaging Surveys\footnote{\url{https://www.legacysurvey.org/dr9/}} program was conducted over more than 14{,}000 deg$^2$ of sky from the Northern hemisphere,  in three optical bands~: $g$ ($470~\rm{nm}$),  $r$ ($623~\rm{nm}$) and $z$ ($913~\rm{nm}$).  The optical surveys were complemented by two infrared bands from the custom ``unWISE'' coadds  \citep[\eg][]{Meisner2017} of the  all-sky data of the Wide-field Infrared Survey Explorer (WISE) satellite \citep{Wright2010}, namely: $W1$ ($3.4~\rm{\mu m}$) and $W2$ ($4.6 ~\rm{\mu m}$).  A full description of the Legacy Imaging Surveys is available in \cite{Dey2019}.
The optical bands were collected via three independent programs:

\begin{itemize}
\item[$\blacktriangleright$] The Beijing-Arizona Sky Survey \citep[BASS;][]{Zou2019} observed $\sim$5{,}100 deg$^2$ of the North Galactic cap (NGC) in $g$ and $r$ using the 2.3-meter Bok telescope. The area surveyed corresponded to approximately $\dec > 32.375~\rm{deg}$.
\vspace{3mm}
\item[$\blacktriangleright$] The Mayall z-band Legacy Survey \citep[MzLS;][]{Silva2016} provided $z$-band observations over the same footprint as BASS using the 4-meter Mayall telescope. Because the median value of the PSF size is significantly better than in the BASS data, the MzLS data are critical for deblending sources and deriving source morphology.
\vspace{3mm}
\item[$\blacktriangleright$] The Dark Energy Camera Legacy Survey (DECaLS) was performed with DECam \citep[the Dark Energy Camera;][]{Flaugher2015} on the 4-meter Blanco telescope. DECaLS observed the bulk of the Legacy Imaging Surveys footprint in $g$,  $r$ and $z$.  DECam was initially built to conduct the Dark Energy Survey (DES) and DECaLS expanded the DES area using publicly available DECam time.  \rc{However, DES imaging is significantly deeper than standard DECaLS observations as it is covered by more exposures (more than 4 in each band).}
\end{itemize}

The median values of the PSF depth and  PSF size that quantify the quality of the photometry,  as explained in Section~\ref{sec:features},  are given in Table~\ref{tab:quality_photometry} for each program.  Figure~\ref{fig:psfdepth_r} shows the PSF Depth $r$ in the Legacy Imaging Surveys and highlights three distinct regions:
\begin{enumerate}
\item In blue,  the combination of BASS and MzLS covering the northern part ($\sim$5{,}100 deg$^2$) of the DESI footprint (designated {\em North} hereafter).
\item In red,  the DES part of DECaLS covering $\sim$4{,}600 deg$^2$ (designated {\em DES}).
\item In green,  the non-DES part of DECaLS covering $\sim$9{,}900 deg$^2$ (designated {\em South}).
\end{enumerate}
The region around the Large Magellanic Cloud ($\ra,\dec$ in the ranges $[52^\circ,  100^\circ]$ and $[-70^\circ,  -50^\circ]$ respectively) is excluded from our study  as it is heavily contaminated by stars.

\subsection{QSO target selection}
\label{sec:selection}

\begin{figure}
	\includegraphics[scale=1]{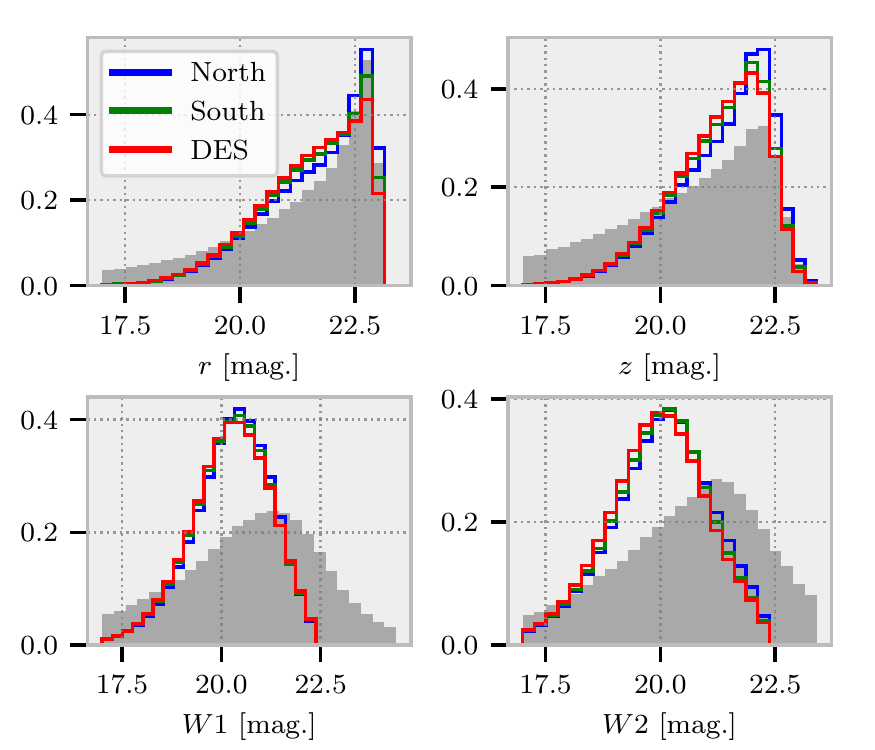}
    \caption{Magnitude distribution of the QSO targets in the $r$,  $z$,  $W1$ and $W2$ bands.  These distributions are shown for the three independent imaging footprints in Figure~\ref{fig:psfdepth_r}.  Blue is for the North,  green for the South and red for DES.  Each grey histogram depicts the magnitude distribution for the parent sample of sources (PSF sources with $r$ < 23) from which QSO targets are selected.}
    \label{fig:Distribution_qso_target.pdf}
\end{figure}

\begin{figure*}
	\includegraphics[scale=1]{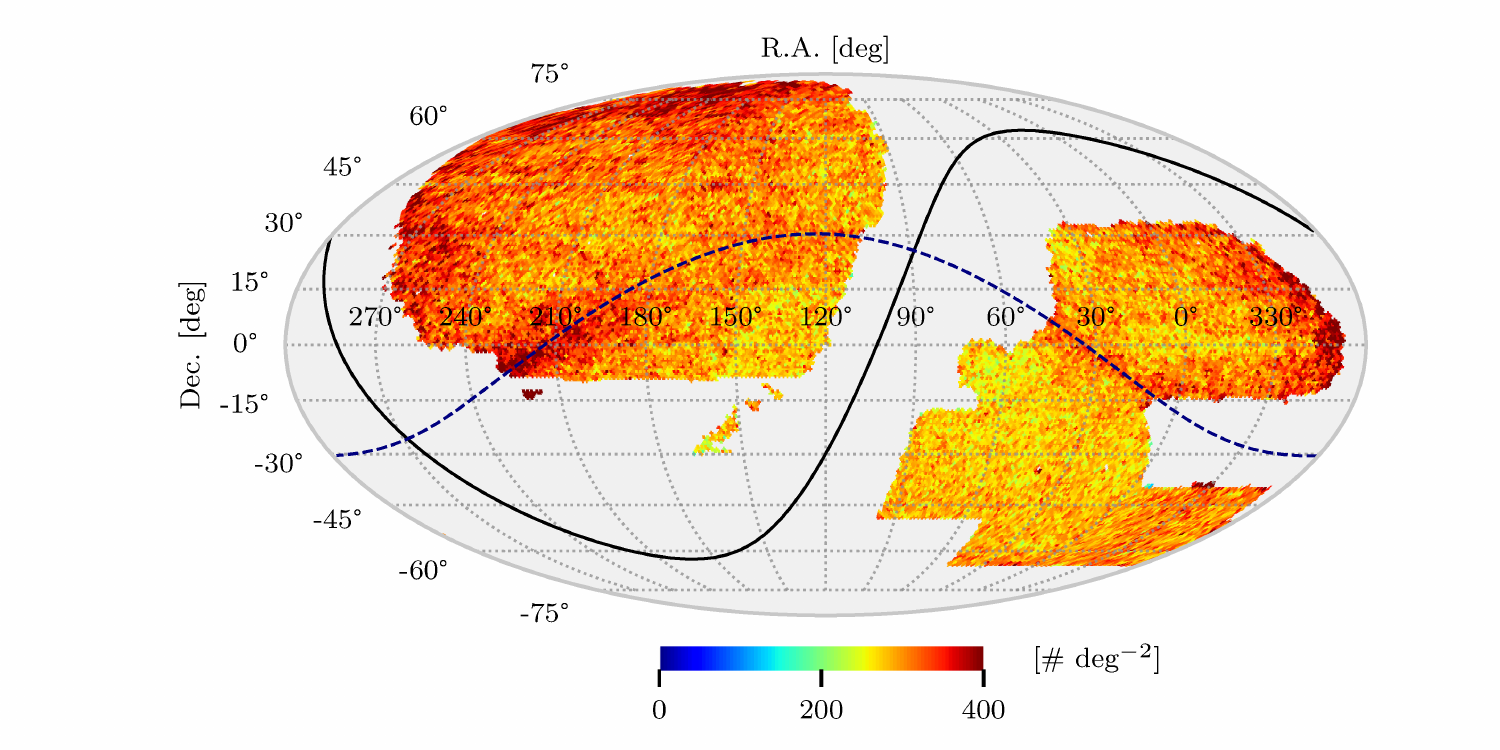}
    \caption{\rc{Density map of the DR9 QSO target selection.  The solid black line indicates the Galactic plane and the blue dashed line indicates the plane of the Sagittarius Stream. }}
    \label{fig:targets_dr9}
\end{figure*}

The  QSO selection  used in this study is an updated version of the preliminary  selection  described in \cite{Yeche2020}. It follows the same general principles but is now based upon  Legacy Surveys DR9 instead of DR8. We briefly summarize below the key properties of the selection.

QSOs are galaxies where a central supermassive black hole accretes large amounts of relativistic gas. The accreting region usually outshines the host galaxy by a large factor while being too small to be resolved, making these distant objects  appear to us as point sources. The QSO selection therefore focuses solely on non-extended sources (\ie sources with point source morphology) in the DR9 catalogue to avoid an almost 10-fold contamination by galaxies. The QSO selection then uses colours to distinguish the QSOs from other point-sources (\ie typically stars).  We restrict the selection to magnitudes in the range $16.5<r<23.0$ to reduce the stellar contamination at the bright end and limit to sources with reliable colours at the faint end.  In addition,  cuts on the WISE bands ($W1 < 22.3$ and $W2 < 22.3$) are used to further limit contamination by stars.  We only consider sources  that are not near bright stars,  galaxies or globular clusters, \ie excluding objects with  {\tt MASKBITS}  1, 12 or 13 set in the Legacy Surveys catalogues\footnote{\url{https://www.legacysurvey.org/dr9/bitmasks/}}.

In the absence of $u$ band,  we cannot apply a selection that relies upon the  UV excess of quasars \citep[e.g.][]{Myers2015}.  In DESI, the main discrimination between quasars and stars is therefore obtained from the fact that quasars are about two magnitudes brighter in the near-infrared at all redshifts than stars of comparable  magnitudes and colours in the optical. The selection method uses a Random Forest classification taking as input the $r$-band magnitude and all colours from two-band combinations of  $grzW1W2$. The selection then identifies QSO candidates as sources that have a probability exceeding an $r$-dependent threshold. To take into account the differences in the photometry,  the probability thresholds are tuned independently over the three imaging footprints in Figure~\ref{fig:psfdepth_r} to produce a median density  of 300 targets per deg$^2$ in \rc{weakly stars contaminated regions.} For the North (\textit{resp.}\ South), the median is computed in box of $\ra,\dec \in [120^\circ, 240^\circ] \times [32.3^\circ, 40^\circ]$ (\textit{resp.}\ $[120^\circ, 240^\circ] \times [24^\circ, 32.2^\circ]$), and over all of the DES footprint.  \rc{Since DES is less contaminated by stars than the other regions,  the target density is on average smaller than 300 targets per deg$^2$. The median in DES is 285 targets per deg$^2$.} The heat map of the resulting  selection is illustrated in Figure~\ref{fig:targets_dr9}.

Figure~\ref{fig:Distribution_qso_target.pdf} shows the magnitude distribution of the QSO targets in the $r$,  $z$,  $W1$ and $W2$ bands and demonstrates that these distributions are similar for each of the imaging footprints highlighted in Figure~\ref{fig:psfdepth_r}.  The three magnitude limits imposed on the selection are clearly visible on the corresponding histograms.  Note that the $r<23.0$ limit also affects the $z$-band distribution,  producing a sharp drop-off at the  faint end for objects in the top-right panel.  The two Gaussian distributions for  $W1$ and $W2$  (bottom panels) demonstrate that the selection of a QSO target is not limited by the depth of the optical imaging (DECaLS or BASS/MzLS) but is sensitive to the determination of the fluxes in the WISE imaging.

To help contextualize which bands guide the DESI QSO target selection,  it is worth noting that the colours that carry the largest weight in the selection are first $z-W2$ and $z-W1$ and then $g-r$,  $W1-W2$ and  $g-z$.

\subsection{Overdensity and contamination}
\label{sec:contamination}
Figure~\ref{fig:targets_dr9} exhibits several  regions with higher  density of QSO targets than average:
\begin{itemize}
	\item[$\blacktriangleright$] The overdensity near the Galactic plane: the stellar density is higher near the Galactic plane (the black line in Figure~\ref{fig:targets_dr9}), which increases the stellar contamination in the QSO target selection.  This effect is not obvious  in the region bounded by $\ra,\dec \in [120^\circ,  140^\circ] \times [-10^\circ,  15^\circ]$,  because the lower $W1/W2$ PSF depth counters the excess of targets caused by the higher stellar density.
	\vspace{3mm}
	\item[$\blacktriangleright$]  The overdensity in the Sagittarius Stream (see also Section~\ref{sec:stars_model}): the stellar population of the Sagittarius Stream (which tracks the blue line in Figure~\ref{fig:targets_dr9}) is different from the Galactic stellar population. Most of the stars in the stream are bluer than Galactic stars and tend to have similar colours to the bulk of QSOs.  Since there is little available data at magnitudes as faint as the stream,  these stars cannot be efficiently rejected and end up contaminating our selected sample of QSO candidates.
	\vspace{3mm}
	\item[$\blacktriangleright$]  The overdensity in the North: the DESI QSO target density increases with declination.  This could be due to the poorer PSF depth in the $z$ band in this region.  This is likely caused by imaging depth decreasing at higher declination due to increasing airmass that was not compensated for by additional exposure time in the MzLS observing strategy.  Since the $z$ band plays a crucial role in the QSO selection,  the discriminating power between stars and DESI QSO targets is reduced at higher declinations.
	\vspace{3mm}
	\item[$\blacktriangleright$] The DES footprint is the least contaminated region,  as expected from the high quality of the photometry in this region.  \rc{Note that DES region is even so contaminated in the region of the crossing of the Sagittarius Stream.}
\end{itemize}

\section{Methodology}
\label{sec:method}
All the density  and feature maps discussed in this section will be pixelized using \texttt{HEALPix}\footnote{\url{http://healpix.sf.net}} \citep{Gorski2005} with $N_{\rm side} = 256$.  All \texttt{HEALPix} operations are done using the \texttt{healpy}\footnote{\url{https://healpy.readthedocs.io/en/latest/}} package \citep{Zonca2019}. The choice of $N_{\rm side}$ (\ie the size of the pixelization) is justified in \ref{sec:256vs512}.

\subsection{Features}
The aim of the systematic mitigation is to correct for spurious density fluctuations in the target selection without suppressing the cosmological signal contained in the targets' clustering. \rc{We restrict our study to ``features'' directly linked to the observational properties such as imaging quality and physical properties that could altered the observations as the stellar density.} We are careful to avoid incorporating parameters that relate to specific positions in the sky.  For instance,  we do not use the  Modified Julian Date (MJD) as a feature \citep[as in, \eg,][]{Rezaie2019}, since the date of an observation directly translates into the position observed on that date.  

\subsubsection{Observational Features from DR9}
\label{sec:features}
\begin{figure*}
	\hspace*{-6mm}
    \includegraphics[scale=1]{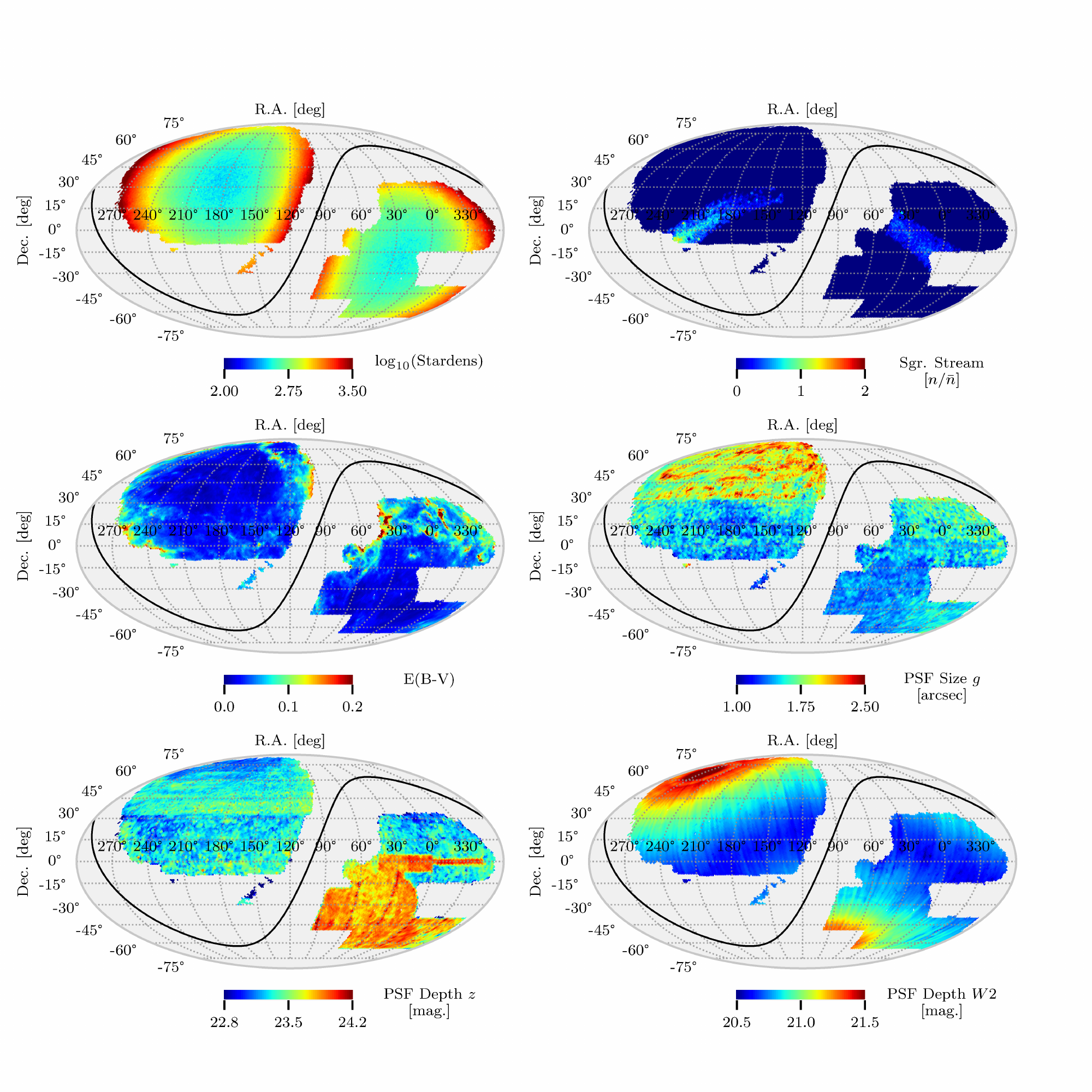}
    \centering
    \caption{Maps of the most important features used in our systematics mitigation. The difference between the three imaging footprints highlighted in Figure~\ref{fig:psfdepth_r} is clearly visible in the {\em PSF Depth z} and {\em PSF Size g} feature maps.}
    \label{fig:feature_map}
\end{figure*}

We use ten observational features to describe the systematic effects in the DESI QSO target selection.  Whether our features are sufficient to describe spurious density fluctuations will be verified after the mitigation procedure by checking the isotropy of the corrected target selection map (see Section~\ref{sec:corrected_targets}).  The impact of mitigating systematics on the observational features themselves will be presented in Section~\ref{sec:systematic_plots}.

The feature maps are generated using the script \texttt{bin/make\_imaging\_weight\_map} from the \texttt{desitarget} package\footnote{\url{https://github.com/desihub/desitarget}}. The maps, which we describe below, are extracted from the random catalogues provided as part of DR9 of the Legacy Imaging Surveys\footnote{\url{https://www.legacysurvey.org/dr9/files}}, with the exception of the stellar density map:
\begin{itemize}
	\item[$\blacktriangleright$]  Stellar density (Stardens) $[\rm{deg}^{-2}]$: Density of point sources from Gaia DR2 \citep{GaiaCollaboration2018} in the magnitude range: $12 < \textrm{PHOT\_G\_MEAN\_MAG} < 17$.  
	\vspace{3mm}
	\item[$\blacktriangleright$]  E(B-V) $[\rm{mag}]$: Galactic extinction from \cite{Schlegel1998} as modified by \cite{Schlafly2011}.
	\vspace{3mm}	
	\item[$\blacktriangleright$]  PSF Depth $[1/\rm{nanomaggies}^2]$ in $r$, $g$, $z$, $W1$, $W2$:  PSF depth is the 5-sigma point-source magnitude depth\footnote{For a $5\sigma$ point source detection limit in band $x$, $5/ \sqrt{x}$ gives the PSF Depth as flux in nanomaggies and $-2.5 \left( \log_{10}(5/ \sqrt{x}) - 9 \right)$ gives the corresponding magnitude (see \url{https://www.legacysurvey.org/dr9/catalogs/}). }.	
The dependence of target selection on  PSF depth is governed by two competing effects.  On the one hand, the number of resolved objects increases with  PSF depth. On the other hand,  the determination of the flux is  better for brighter objects,  which means that contaminants are more easily rejected,  resulting in a reduced number of targets.  In this study,  the $z$ depth does not limit the target selection (cf.\ Section~\ref{sec:selection}): it only affects the measurement of the $z$ flux. Therefore, the target density decreases with increased $z$ PSF Depth.  In contrast,  fluxes in $W1$ and $W2$ are obtained with forced-photometry.  This allows fluxes to be measured for fainter objects that are detected at marginal significance in the WISE imaging \citep[see][]{Dey2019}. Because such fluxes are noisy, the resulting colours scatter in regions of insufficient depth.  Corresponding objects are rejected by the target selection not because they are not quasars but because their colours are not in the region populated by QSOs.  Conversely, when the PSF Depth in $W1$ or $W2$ is higher, the colours are more precise,  and a greater number of targets are selected as genuine QSOs.
	\vspace{3mm}
	\item[$\blacktriangleright$]  PSF Size $[\rm{arcsec}]$ in $r$, $g$, $z$: Inverse-noise-weighted average of the full width at half maximum of the point spread function, also called the delivered image quality.  A small value of PSF size corresponds to a good image resolution,  which leads to more precise fluxes and improved target selection. 
\end{itemize}

Figure~\ref{fig:feature_map} shows 5 of the 10 observational feature maps from DR9.  
It is important to note that these 10 features are correlated, and the level of correlation depends on the imaging footprint (North, South, DES).  Stellar density and E(B-V) are positively correlated across the entire sky but E(B-V) contains additional information. This can be seen by comparing the top two \rc{left} panels in Figure \ref{fig:feature_map} in the region near $(\ra, \dec) = (0^\circ, 15^\circ)$.  The $W1$ and $W2$ PSF Depths are highly correlated in all of the three imaging footprints of interest.  However,  we include both as our analysis will show that the $W2$ PSF Depth is particularly heavily weighted by the Random Forest in the North, indicating that this feature has extra information which is not contained in the $W1$ PSF Depth.  For the optical observational features,  two cases are of particular note.  In the North,  the $g$ and $r$ bands were collected by the same camera and so their observational features are positively correlated, but these features are not correlated with the $z$-band, which was independently obtained by MzLS.  In the South and in DES,  the three bands were collected with the same camera and therefore are all positively correlated.  Finally,  the $W1/W2$ PSF Depths are more correlated with the other features in the DES region, as compared to the North and the South, since all the feature maps are more uniform in the DES footprint.

\subsubsection{Sagittarius Stream Model}
\label{sec:stars_model}
The target density map (Figure~\ref{fig:targets_dr9}) shows a significant excess in the Sagittarius Stream region.  The Sagittarius galaxy is one of the closest galaxies orbiting around the Milky Way.  The gravitational forces create two tidal arms called streams wrapping the Milky Way with the same orbit~\citep{Newberg2001,Majewski2003}. 

The Sagittarius contamination (see the blue dashed line in Figure~\ref{fig:targets_dr9}) occurs mainly in the South footprint but also touches the DES footprint.  However,  none of the feature maps discussed so far contain a pattern matching this contamination. Thus, an additional feature is required in our analysis.  To model this feature, we use the Sagittarius Stream catalogue derived in \cite{Antoja2020}. This catalogue is built from the Gaia DR2 catalogue, identifying stars in the Stream via their proper motions. Matching the \citeauthor{Antoja2020} catalogue to the SDSS DR16 QSO catalogue \citep{Lyke2020} on position, we find that some of the stars are actually known QSOs. To generate the Stream feature  shown in Figure~\ref{fig:feature_map} (top right panel), we remove the known QSOs and apply an $r>18$ cut to only select faint stars. A fainter cut would better match the QSO selection, but the \citeauthor{Antoja2020} catalogue does not contain a sufficient number of objects faint in $r$-mag to apply a fainter cut. 

\subsubsection{Three imaging footprints}
\begin{figure}
    \includegraphics[scale=1]{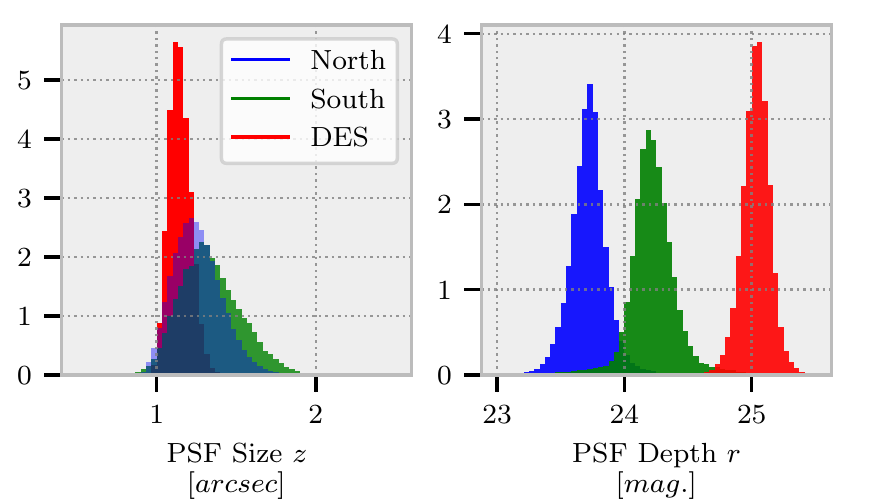}
    \centering
    \caption{Density distribution of the PSF Size $z$ and PSF Depth $r$ for the three imaging footprints shown in Figure~\ref{fig:psfdepth_r}.  These two histograms show that a joint analysis on the three footprints is difficult \rc{since some features share the same support and others do not.}}
    \label{fig:feature_hist}
\end{figure}

The three footprints of the Legacy Imaging Surveys (North,  South,  DES; as defined in Section~\ref{sec:DR9}),  exhibit distinct imaging properties.  As shown in Figure~\ref{fig:feature_hist}, while the PSF size in the $z$ band is similar in the North and South, the $r$-band depth distributions are very different in the North, South and DES footprints. For instance,  a PSF depth of $24.7$ in $r$ leads to a small overall target overdensity in DES (cf.\  Figure~\ref{fig:systematics_plot_des}) while it corresponds to an underdensity in the South (cf.\  Figure~\ref{fig:systematics_plot_south}).  The imaging properties in the South footprint are similar in the North (NGC) and  South (SGC) Galactic Cap. There is therefore no reason to split the South footprint in two.

In addition,  the selection threshold in the QSO target selection is set independently in each of these three footprints.  We therefore model systematic effects in the three footprints independently.

\subsection{Systematic Model}
We will identify each position on the sky with the corresponding pixel number $i$ from a \texttt{HEALPix} pixelization.  For each imaging footprint,  we only consider pixels that contain at least one object from the random catalogues. The density of selected targets inside a pixel $i$ is noted $N_i$.  It is derived from the observational target density as
\begin{equation}
N^{\rm obs}_i = f_i  \times N_i, 
\end{equation}
where $f_i$ is the observed fractional area of  pixel $i$ calculated as the number of random targets inside the pixel divided by the nominal density of randoms,  and $N_i$ is the true number of quasars in the pixel in the absence of any systematic effects. $N_i$ is related to the mean target density $N_0$ by
\begin{equation}
N_i= N_0\left(1 + \delta_i \right),
\end{equation}
where $\delta_i$ is the overdensity that contains the cosmological information. 

$N_0$ is unknown and has to be estimated from the target selection as the mean of the pixel density over the footprint weighted by $f$: $\hat{N_0} = \langle N_i \rangle_{i} =  \langle f_i^{-1} \times N^{\rm obs}_i \rangle_{i} $.  Since the target selection is contaminated by stars,  we choose different regions known to be less contaminated to perform this estimation in the three footprints.  We use the same region used to tune the probability selection during the quasar target selection, which was a box of $\ra,\dec \in [120^\circ, 240^\circ] \times [32.2^\circ, 40^\circ]$ for the North footprint, a box of   $\ra,\dec \in [120^\circ, 240^\circ] \times [24^\circ,  32.3^\circ]$ for the South footprint and the entire footprint for DES.  Note that even these regions are contaminated by stars, such that the actual QSO target density is smaller.

Systematic effects will be taken into account using an additional term $F_i$ such that 
\begin{equation}
N_i = N_0\left(1 + \delta_i \right) \times F_i.
\end{equation}

The aim of the imaging systematics mitigation is to describe $F_i$ as a function of a set of observational features. These features must not depend on the sky position, to avoid suppressing the cosmological signal.   We assume that the imaging systematics can be completely explained by our set of observational features.  This assumption can be validated by the uniformity of the target selection density map after  mitigation (cf.\, Section~\ref{sec:corrected_targets}). 

The features in the pixel $i$ are denoted $s_i$ which is an $n$-dimensional vector where $n$ is the number of features ($n=11$ in our case: 10 observational features from Section~\ref{sec:features} and the Sagittarius Stream feature from Section~\ref{sec:stars_model}). $F$ should depend only on these features and not on the pixel number. We thus rewrite  $F_i$ as  $F(s_i)$, and  $N$ now also depends on the observational features: 
\begin{equation}
N_i(s_i) = N_0\left(1 + \delta_i \right) \times F(s_i).
\end{equation}

We denote $A(S)$ to be the subset of pixel numbers which belong to the same region $S$ in the space of the feature maps.  Averaging over many pixels will suppress the density contrast: $\langle \delta_{i} \rangle_{i \in A(S)} = 0$.  Thus,  if $S$ is sufficiently large, the contamination signal is given by
\begin{equation}
F(S) = \dfrac{\langle N_i(S) \rangle_{i \in A(S)}}{N_0}.
\label{eqn:F_S}
\end{equation}

\noindent This averaging is controlled by hyper parameters of the different regression methods.  For instance in the particular case of the Random Forest (see Section~\ref{sec:rf}),  the averaging is controlled by the minimum number of objects in a leaf.

The systematic correction will be modelled by a weight to apply in each pixel,  defined as 
\begin{equation}
w_i^{\rm sys} = \dfrac{1}{F(s_i)}.
\end{equation}
The correction is only efficient on scales at least as large as the typical size of the pixel at $N_{\rm side}=256$, \ie about $\theta = 0.22 ~\rm{deg}$.  Hence,  the correction is constant within  each pixel.

The regression is performed using only reliable pixels, which we choose to be pixels with $ f_i > 0.9 $, and is then applied to all pixels. This criterion removes pixels that contain too few targets, which could bias the regression.  The pixels that are excluded represent only about 3.8\% of the DR9 footprint, lying mainly at the edges of the footprint and in the region south of $\dec<-10^\circ$ in the NGC.  For studies with a smaller pixel size,  e.g.\ $N_{\rm side}=512$,  the nominal density of the randoms would have to be increased to limit the Poisson noise when determining $f_{i}$.

\subsection{Regression Methods}
We test different methods, utilizing the same feature set, to perform the regression presented above. The initial correction, obtained with a linear regression, turns out to be insufficient (as illustrated in Figure~\ref{fig:systematics_plot_north} -- \ref{fig:systematics_plot_des}), necessitating non-linear regression approaches.  We therefore test two classical machine-learning methods based on \texttt{scikit-learn} \citep{Pedregosa2011}, namely, the Random Forest and Neural Network methods.

\subsubsection{Linear}
The contamination function $F$ is described as a linear function of the observational features: $F(s_i) = a_0 + \sum_{j=1}^{11} a_j s_{ij}$.  The coefficients $a_j$ will be estimated with a least square minimization using the \texttt{iminuit} package \citep{iminuit}.  The $\chi^2$ used for the minimization is defined as 
\begin{equation}
\chi^2 = \sum_i \dfrac{1}{N_i} \left(F(s_i) -\dfrac{N_i}{N_0}\right)^2 + c_{\rm reg} \times \left( \langle F(s_i) \rangle_i  - 1 \right)^2,
\end{equation}
 where $\sqrt{N_i}$ is an estimate of the error for the object count inside a pixel and $c_{\rm reg}$ is a penalty term to regularize the regression.  Since the distribution of ${N_i}/{N_0} $ is not symmetric around 1, the higher number of pixels with ${N_i}/{N_0} < 1$ forces the contamination function to not be centered around 1.  We therefore use a penalty term to flatten the density map around the chosen mean density.  The value of the penalty term depends on the number of pixels used to build $\chi^2$.  In our configuration,  we use $c_{\rm reg} = 2e6$ and we check that as long as $c_{\rm reg}$ is sufficiently large,  its value does not change the result of the regression.
 

\subsubsection{K-fold training}
\label{sec:k-fold}
Machine learning methods tend to overfit the data and have to be used carefully.  Since we cannot create a training sample independent of the data set,  we have to use a K-fold training to avoid over-training.  

A K-fold training is a method that splits the data into K folds.  For each fold,  the method is trained with the remaining K-1 folds then the regression is performed on the isolated fold.  This method guarantees that no data used for the regression is used for the training.  A small value of K ensures no overfitting but the mitigation will be inefficient since the training set would not contain enough information.  A high value of K ensures an efficient mitigation but is more prone to overfitting.  In our case,  we choose to work with 6 folds in the North and the DES footprint and with 12 folds in the South.  Each fold has an area of about 830 $\rm{deg}^2$.

The locations of the folds have to be carefully chosen, since the contamination has distinct causes.  For example,  if all the borders of the Galactic plane in the South footprint were to be put in the same fold,  the machine learning algorithm would not be able to explain related contamination with the K-1 remaining folds, since all the relevant information would have been removed from the training set.  Therefore, we construct folds from small patches of the sky to spread the information across all folds. Such folds can be constructed easily in the \texttt{HEALPix} ``nested'' scheme using the \texttt{scikit-learn} function {\tt GroupKFold}.  The size of the patch matters, as patches that are too small lead to overfitting since all the information is present in all the folds.  We use patches (groups in  \texttt{scikit-learn} language) of 1000 pixels,  which corresponds to an area of $\sim$52\,$\rm{deg}^2$ for each patch. 

\begin{figure}
    \includegraphics[scale=1]{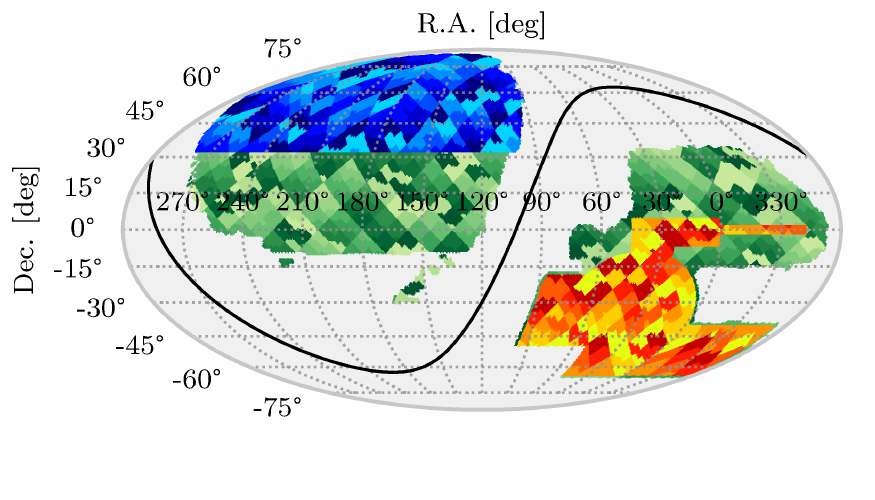}
    \centering
    \caption{Distribution of folds across the three imaging footprints. There are 6 folds in the North (blue region), 12 in the South (green region) and 6 in DES (red region).  Folds were split into small patches so the specific effects that contaminate target selection are always spread across several folds.  The area of each patch is $\sim$52\,$\rm{deg}^2$.}
    \label{fig:fold}
\end{figure}

The patch distribution is shown in Figure~\ref{fig:fold}, which demonstrates that no particular region is described by a single colour.  We checked that the estimated weight remains stable if we slightly vary the number of folds or the size of the patch, and used mocks (cf.\  Appendix~\ref{sec:mocks}) to ensure that no overfitting occurs.

\subsubsection{Random Forest (RF)}
\label{sec:rf}
The first machine learning method used in our regression analysis is the well-known Random Forest algorithm.  It is  easy to parameterise and gives  a helpful classification of the features as a  function of their importance during the regression.  

For the regression, we use the same set of hyper-parameters for each footprint and we do not normalise the data set.  We choose a forest of 200 trees and we  fix the minimum number of samples contained inside a leaf at 20.  This means the average to estimate $F(S)$ (cf.\ Eq.\,\eqref{eqn:F_S}) is computed with at least 20 pixels.  The mean number of pixels in each leaf is 80.  We checked that the minimum sample size in a leaf has no strong impact on the regression during the K-fold training. 

\subsubsection{Neural Network (NN)}
\label{sec:NN}
We also use the Multi-layer Perceptron (MLP) algorithm,  a fully connected type of Neural Network.  Finding the best hyper-parameters for a neural network is difficult. We base our choice on the result of a grid search performed on simulated QSO samples (\ie mocks).  The mocks used for this study are described in Appendix~\ref{sec:mocks}.

For the regression, we use the same set of hyper-parameters for each footprint, but when dedicated mocks for DESI become available, different hyper-parameterisations will be possible.  We use an MLP with 2 hidden layers, comprising 10 neurons for the first layer and 8 for the second (\ie a $(10, 8)$ formalism).  The data are normalised on each fold dividing each feature by the standard deviation estimated in the considered fold.  We hyper-parameterise the MLP with a sigmoid activation function,  a batch size of 1000 and a tolerance of 1e-5.  We use the Adam solver \citep{Kingma} to perform the minimization during the training. 

\section{Systematics Mitigation}
\label{sec:mitigation}

We apply the method presented in Section~\ref{sec:method} to correct for observational systematics in the DESI QSO target selection. We first describe the origin of these systematics  and explain the role of the most important features. We then present the target density after it has been corrected  by our systematic mitigation method. 

\subsection{Systematic Plots}
\label{sec:systematic_plots}

\begin{figure*}
    \hspace*{-6mm}
    \includegraphics[scale=1]{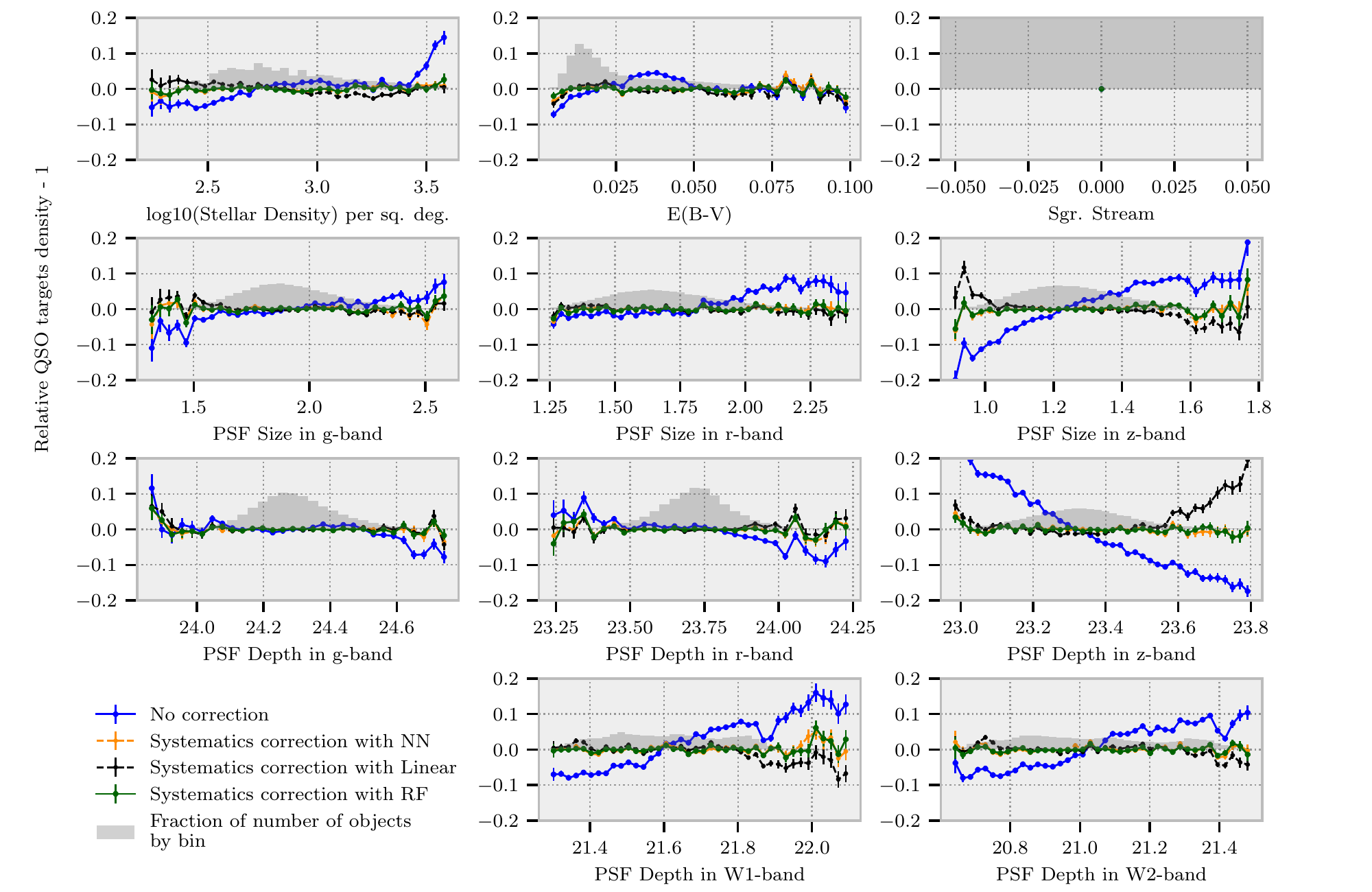}
    \centering
    \caption{Relative QSO target density in the North footprint as a function of each observational feature.  The relative QSO target density is a mean value after rejecting outliers \rc{\ie pixels with a coverage lower than 90\%}.  The blue lines depict the raw DESI QSO target selection. The green (\textit{resp.} yellow / black) lines depict the QSO target selection after correcting for systematic effects using the RF (\textit{resp.} NN / Linear) regression.  The histogram represents the fraction of objects in each bin for each observational feature and the error bars are the estimated standard deviation \rc{of the normalized target density} in each bin.  The three methods all successfully flatten the relative QSO target density as a function of each observational feature.  However,  the linear method is less efficient than the other two methods.}
    \label{fig:systematics_plot_north}
\end{figure*}

\begin{figure*}
    \hspace*{-6mm}
    \includegraphics[scale=1]{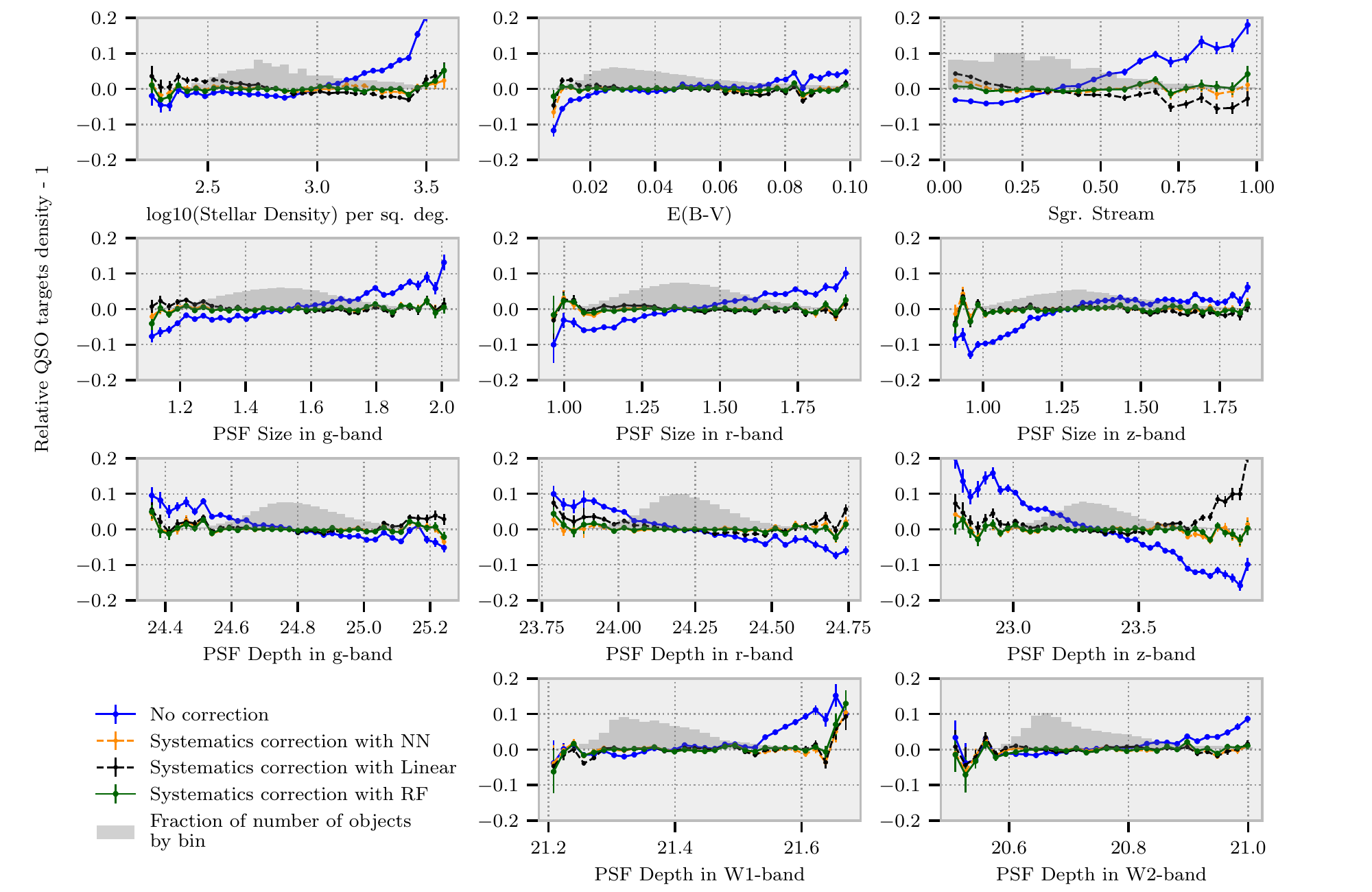}
    \centering
    \caption{Same as Figure~\ref{fig:systematics_plot_north} but for the South imaging footprint.}
    \label{fig:systematics_plot_south}
\end{figure*}

\begin{figure*}
    \hspace*{-6mm}
    \includegraphics[scale=1]{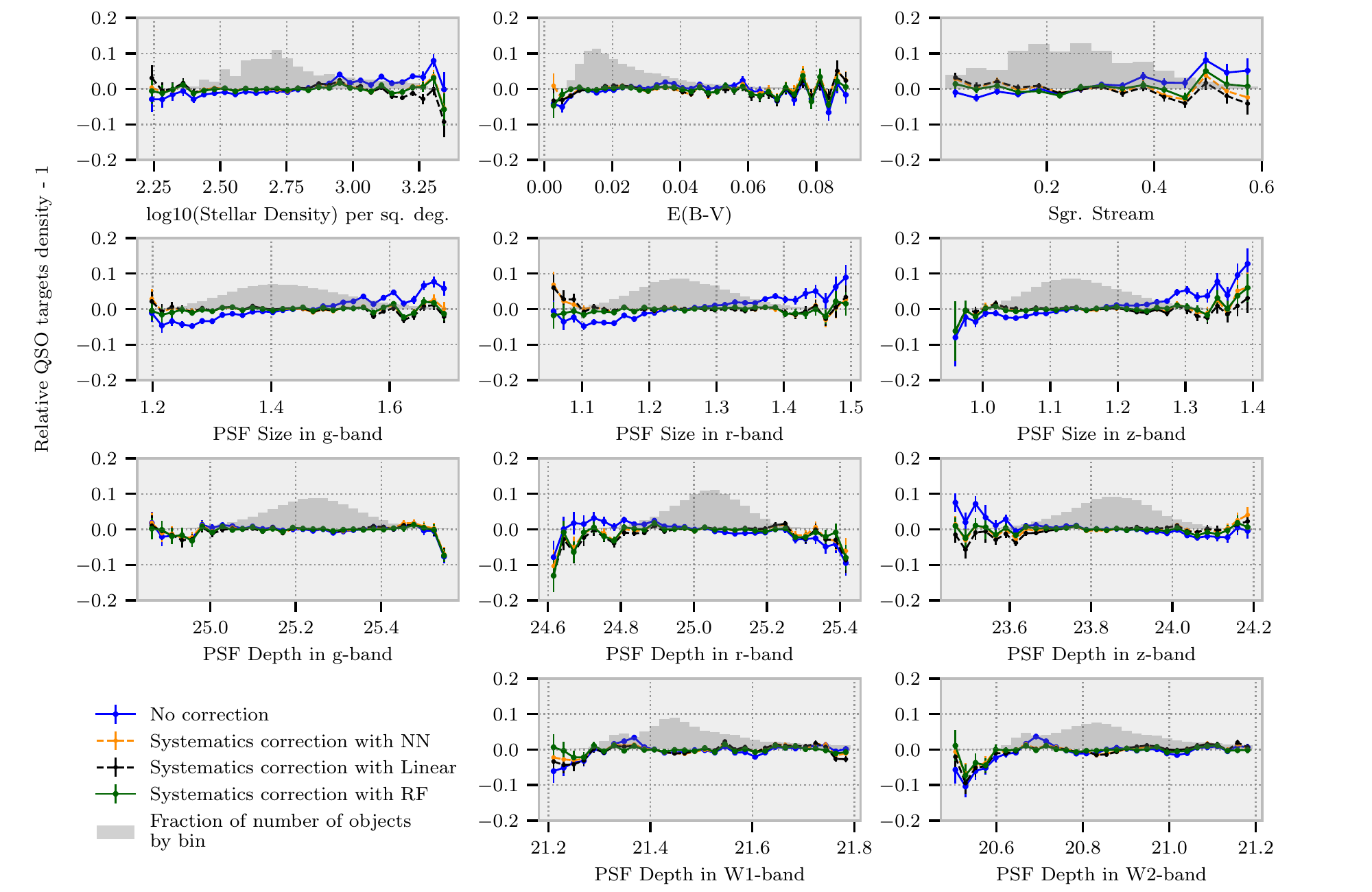}
    \centering
    \caption{Same  as Figure~\ref{fig:systematics_plot_north} but for the DES imaging footprint.}
    \label{fig:systematics_plot_des}
\end{figure*}

To illustrate our method, we plot the relative QSO target density as a function of each observational feature.  We will refer to these plots as ``systematic plots''. In our systematic plots, we center the relative density around 0. The goal of the  correction is to obtain a  relative density that is independent of the value of the observational feature. We produce systematic plots for the North (see Figure\,\ref{fig:systematics_plot_north}), the South (see Figure~\ref{fig:systematics_plot_south}) and for DES (see Figure~\ref{fig:systematics_plot_des}). 

Below,  we give a brief description of the systematic plots in each footprint:
\begin{itemize}
\item[$\blacktriangleright$] North (Figure\,\ref{fig:systematics_plot_north}): $z$ and $W1/W2$ are crucial for the DESI QSO target selection. The plot of the relative density as a function of the PSF Depth $z$ or the PSF Size $z$ shows that when the $z$ observational feature values are for bad observational conditions (small PSF Depth and high PSF Size),  the discriminating power of the target selection algorithm is poorer and  the relative density higher.  In addition to the importance of the $z$ band for the target selection,  the $z$ band benefits from better image quality (smaller PSF Size and higher PSF Depth) than $g$ or $r$. The fluctuation of the relative density as a function of the $g/r$ observational features are therefore weaker but they follow the same general pattern.
As explained in Section~\ref{sec:features}, the impact of the $W1/W2$  PSF Depths differs from the $g/r/z$ depths since the depth of the WISE colours are crucial for selecting QSOs.  Hence,  the number of targets increases with $W1/W2$ PSF Depth.  In addition, the $z$ PSF Depth is smaller in regions where the $W1/W2$ PSF Depth is larger, as shown in the bottom panels of Figure~\ref{fig:feature_map}.  The combined effect increases the target density in this region.

\vspace{3mm}
\item[$\blacktriangleright$] South (Figure\,\ref{fig:systematics_plot_south}): the plots as a function of the $z$ and $W1/W2$ PSF depths at high values exhibit similar behaviour as in the North.  However,  the effect is less significant since it is mainly visible in the $\ra,\dec \in [210^\circ, 270^\circ] \times [15^\circ, 30^\circ]$ region.
In comparison to the North,  the plots as a function of the $g/r/z$ features all show similar trends to each other since these bands were collected with the same camera. The $z$-band dependence is the strongest,  as expected since it is one of the most important bands for the QSO selection.  
The plots as a function of the $W1/W2$ PSF Depth exhibit seemingly unexpected behaviour at low values of the PSF Depth, where the relative density is almost flat.  As explained in Section~\ref{sec:contamination},  the excess of stars near the Galactic plane or the Sagittarius Stream counteracts any expected decrease of targets due to lower values of the $W1/W2$ PSF Depth (cf., in particular,  the $\ra,\dec \in [120^\circ,  150^\circ] \times [-10^\circ,  15^\circ]$ region).
The plots in the top panel of Figure\,\ref{fig:systematics_plot_south} (stellar density,   dust and Sagittarius Stream) indicate a higher relative density due to the presence of stars. These features explain the stellar contamination near the Galactic plane and inside the Sagittarius Stream.  Some stars have similar colours to QSOs and are therefore selected as QSO targets.  More stars thus enter the QSO selection in regions of higher stellar density.

\vspace{3mm}
\item[$\blacktriangleright$] DES (Figure\,\ref{fig:systematics_plot_des}): The fluctuations of the relative density are much lower than in the two other footprints. The observational features for $g/r/z$, and especially for the $z$ band, are  better in DES than in both the South and the North (cf.\ the median value of these features in Table~\ref{tab:quality_photometry} and the distribution of the PSF Depth Size $z$ and the PSF Depth $r$ in Figure~\ref{fig:feature_hist}).  This results in excellent discrimination of QSOs from stars in the target selection process. So,  DES is the least contaminated region and exhibits smaller fluctuations in relative density as a function of the observational features.  
\end{itemize}

In Figures~\ref{fig:systematics_plot_north} -- \ref{fig:systematics_plot_des}, we also plot the three different regression methods we apply for each footprint: RF in green,  NN in yellow and Linear in black.  These plots show that the mitigation works well and correctly flattens  each feature.  The difference in efficiency between the Linear and the non-linear methods is particularly obvious for the North and South footprints, where the contamination is stronger.  For example,  the linear correction fails when the relative density as a function of the PSF depth $z$ is large, as shown in Figure~\ref{fig:systematics_plot_north}.  This illustrates the non-linearity of the contamination and justifies the use of the machine learning methods. Both the RF and the NN perform well in correcting the non-linear systematics, although small differences can be found between these two methods.  A more detailed comparison between the RF and NN will be done in Section~\ref{sec:method_comparaison}.

\subsection{Importance Features}
\label{sec:importance_feature}
The Random Forest algorithm includes a specific tool called {\em importance features}. Importance is a measure of which features most affect the regression.  The importance features metric for the RF implemented in \texttt{scikit-learn} is based on the mean decrease in impurity (MDI), which is also called the Gini importance.  The metric corresponds to averaging (over all the trees of the forest) the impurity reduction of each node weighted by the ratio of the training set passing through each node.  The {\em permutation importance} was also tested and yields similar result in our data set than the Gini importance. 

A feature with a higher importance value is more discriminating than a feature with a lower importance value, so the importance values are useful to identify  the observational features that lead to contamination, to first order. It is worth noting that a feature with a low importance value is not necessarily useless and can still improve the training compared to a case where we remove it.  For instance,  the Sagittarius Stream feature is necessary to correct the over-density in the Sagittarius region but it is not a high-value importance feature since it is useful only for a small number of pixels that vary in a manner that is quite different from other pixels in the footprint.  This well-known bias can be circumvented using the {\em permutation importance}, where the Sagittarius Stream feature is ranked as one of the most important in the South. 

\begin{figure}
    \includegraphics[scale=1]{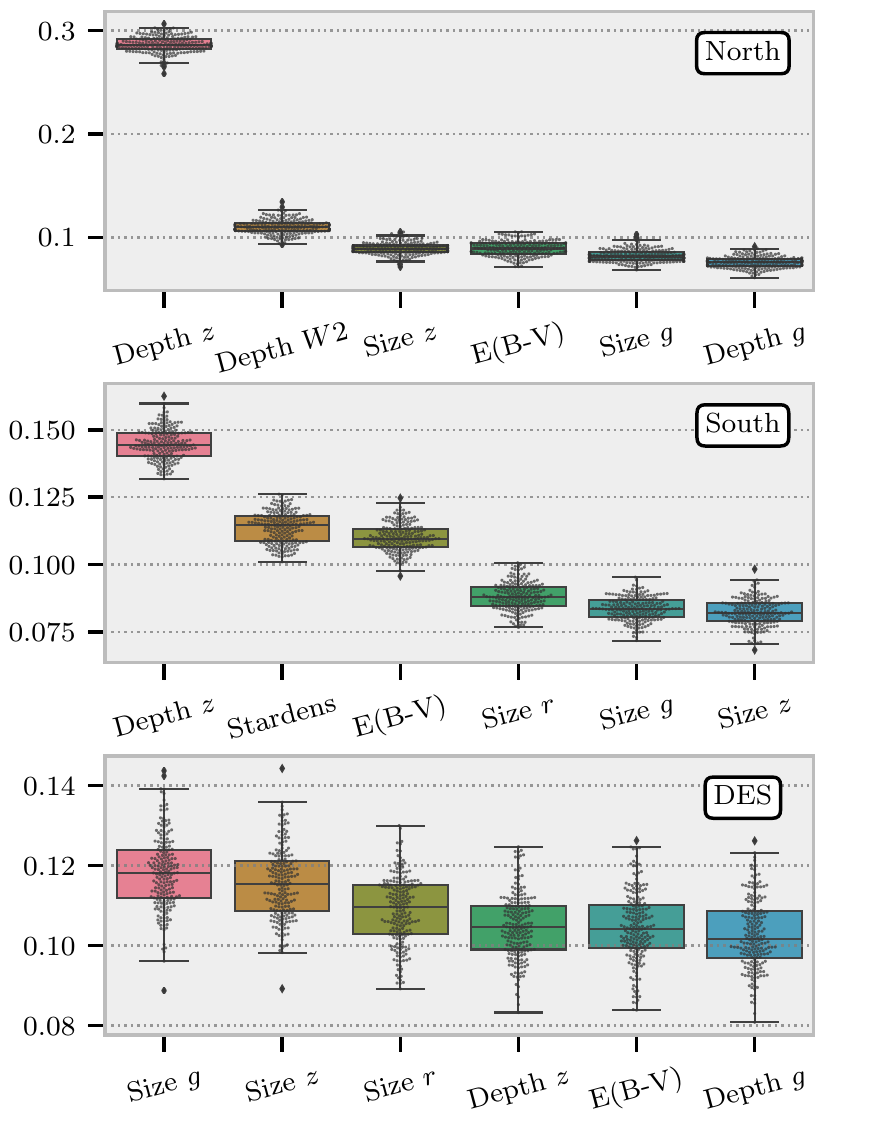}
    \centering
    \caption{Feature importance calculated using the Random Forest method.  Only the 6 most important features are plotted for each footprint.  Each dot represents the feature importance value found with one decision tree.  The box plot is assembled across all the trees of the forest.  The values of the importance's cannot be compared between the three regions.}
    \label{fig:feature_importance}
\end{figure}

In Figure~\ref{fig:feature_importance}, we plot the importance for each feature.  We only plot the six most important features for each footprint, since the other features have about the same value as the sixth most important feature. 
We recover the expected most important features  described in Section~\ref{sec:systematic_plots}:

\begin{itemize}
\item[$\blacktriangleright$] North: The importance feature analysis makes it clear that the PSF Depth $z$ and the PSF Depth $W2$ play a key role in the contamination of the target selection in the North.
\vspace{3mm}
\item[$\blacktriangleright$] South: The South region is strongly contaminated by stars from the Galactic plane as highlighted by the importance of the stardens and EBV features.  The Sagittarius  Stream feature has a low importance even if it is crucial to correct the Sagittarius Stream region (as explained above).  The PSF Depth $z$ plays an important role in this overdensity, as already noted.  Surprisingly,  the $W2$ PSF Depth does not appear as an importance feature, even though it explains the under-density near the  anti-Galactic pole.
\vspace{3mm}
\item[$\blacktriangleright$] DES: No clear importance feature emerges, as  expected given that the DES region is the least contaminated, most uniform, region.
\end{itemize}

\subsection{Quasar target selection after correction}
\label{sec:corrected_targets}

\begin{figure}
    \includegraphics[scale=1]{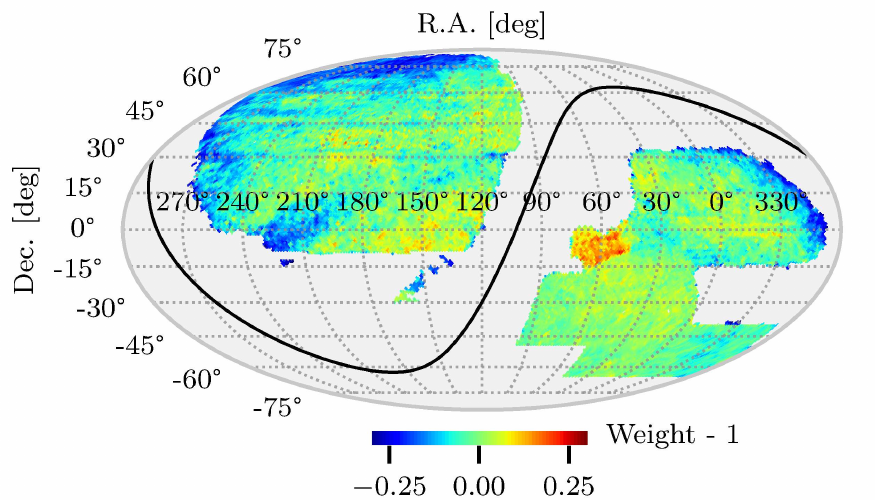}
    \centering
    \caption{Distribution of the systematic correction weights \rc{(weight - 1)} across the Legacy Imaging Surveys footprint.}
    \label{fig:weight}
\end{figure}

\begin{figure}
    \includegraphics[scale=1]{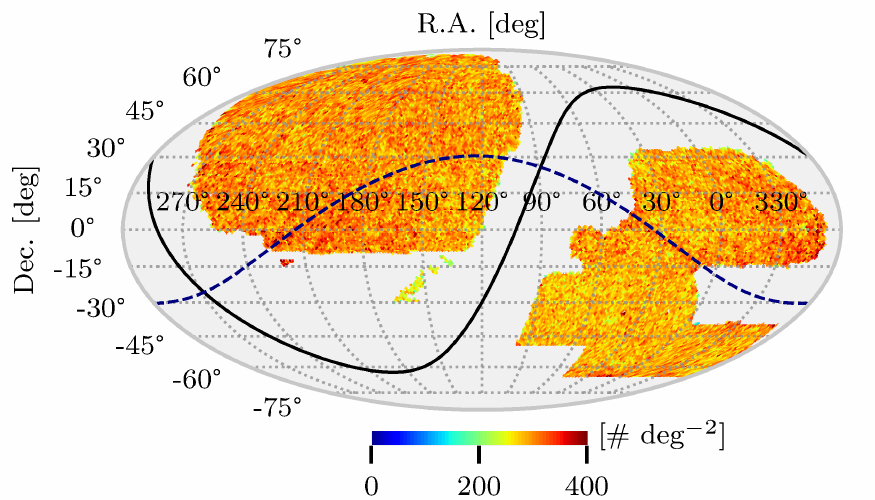}
    \centering
    \caption{Density map of the QSO target selection after mitigating systematics using the RF method. The density map is quasi-uniform compared to the initial density map shown in Figure~\ref{fig:targets_dr9}. The solid black line depicts the Galactic plane and the blue dashed line depicts the plane of the Sagittarius Stream.}
    \label{fig:targets_corrected}
\end{figure}

We construct a pixel weight map to correct the over or under densities of the target selection.  Figure~\ref{fig:weight} shows the weight map obtained with the RF regression.  We then multiply the density of the DESI QSO target selection by the weight on a pixel-by-pixel basis.  Figure~\ref{fig:targets_corrected} shows the corrected QSO  selection.  The map is almost completely uniform at this resolution.  The largest overdensities vanish after mitigation, which confirms that the set of observational features that we considered suffices to explain most of the observed large-scale target density variations.

\begin{figure}
    \includegraphics[scale=1]{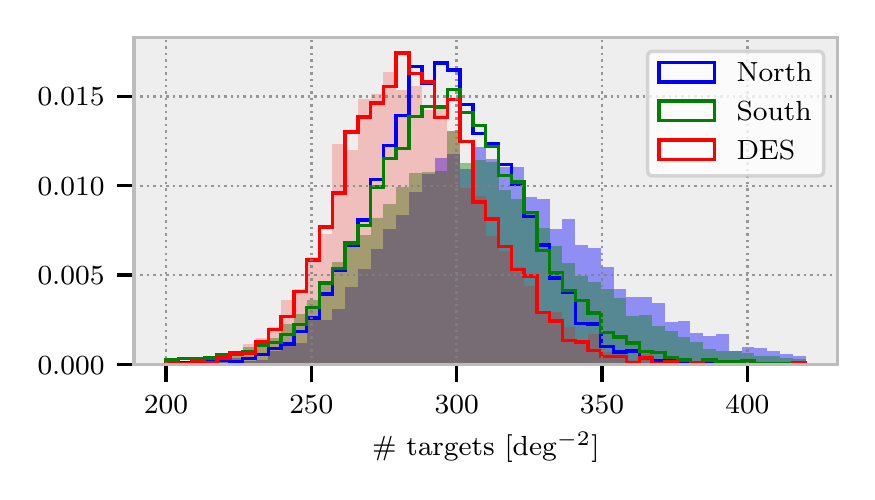}
    \centering
    \caption{\rc{Density distribution of the QSO target selection for non corrected (full histogram) and corrected (line) cases in the three photometric footprints.  After the correction,  the width of the histogram is smaller in each photometric footprint.  As mentioned in Section~\ref{sec:selection}, the density in DES is lower than in the North or in the South.  }}
    \label{fig:pix_distribution}
\end{figure}

\rc{The density distribution of the QSO target selection is shown in Figure~\ref{fig:pix_distribution}. The full histograms are before applying the systematic mitigation and the lines are for after.  The systematic mitigation acts on the width of the histograms removing the over or under-density pixels.  This effect is smaller in DES than in the North and in the South which confirms the visual inspection of the Figure~\ref{fig:targets_dr9} and \ref{fig:targets_corrected}. The difference in mean density between DES and the two other photometric footprints was mentioned in Section~\ref{sec:selection}}

\section{Angular Correlation and Clustering}
\label{sec:clustering}
Our systematic mitigation method is able to correct fluctuations in density as a function of our observational features. In this section, we measure the impact of the systematic mitigation on the angular correlation function.

\subsection{2-point Correlation function}
\subsubsection{Definition}
The 3-dimensional 2-point correlation function $\xi(\mathbf{r})$ describes the excess probability to find a pair of objects inside two infinitesimal volumes $dV_1$ and $dV_2$ separated by $\mathbf{r}$: 
\begin{equation}
dP(\mathbf{r}) = \bar{n}^2 \left( 1 + \xi(\mathbf{r}) \right) dV_1 dV_2, 
\end{equation}
where $\bar{n}$ is the mean density.  It is related to the contrast density by $\xi(\mathbf{r}) = \langle \delta(\mathbf{x}) \delta(\mathbf{x + r}) \rangle_{\mathbf{x}}$.  The cosmological principle ensures that $\xi$ depends only on $r$.

The same definition can be extended to the 2-dimensional case \ie the angular correlation where volumes are replaced by solid angles and distances by angular distances: 
\begin{equation}
dP(\theta) = n_0^2 (1 + \omega(\theta)) d\Omega_1 d\Omega_2, 
\end{equation}
where $n_0$ is the mean angular density. 

\subsubsection{2-point correlation function estimator}
We estimate the angular correlation function $w(\theta)$ using the Landy-Szalay estimator derived in \cite{Landy1993}: 
\begin{equation}
\hat{w}(\theta) = \dfrac{aDD - bDR + RR}{RR},
\end{equation}
 where $DD$,  $DR$,  $RR$ refer to the weighted pair counts $data-data$,  $data-random$ and $random-random$ with an angular separation $\theta$.  The normalization terms are
 \begin{equation}
a = \dfrac{\sum  \limits_{i \neq j} w_i^Rw_j^R}{\sum \limits_{i \neq j} w_i^D w_j^D} \text{\hspace{0.4cm} and  \hspace{0.4cm}} b = \dfrac{\sum  \limits_{i \neq j} w_i^Rw_j^R}{\sum  \limits_i w_i^D \sum  \limits_j w_j^R},   
 \end{equation}
where $w_i^D$ (\textit{resp.} $w_i^R$) is the weight for the data (\textit{resp.} random).
The Landy-Szalay (LS) estimator is known to be unbiased and to have minimal variance in the limit of an infinitely large random catalogue with a volume greater than the scales considered, and for weak correlations ($w(\theta) << 1$). 

We use the package \texttt{CUTE}\footnote{\url{https://github.com/damonge/CUTE}} \citep{Alonso2012} to perform the estimation.  \texttt{CUTE} is a fast implementation written in C and using OpenMP and MPI.

\subsection{Angular correlation of the DR9 quasar target selection}
\rc{The systematic mitigation method is checked to avoid over-fitting.  This is done in Appendix~\ref{sec:mocks} by applying the mitigation method to contaminated mocks.  Then, we compute the angular correlation of the DR9 quasar target selection corrected by the systematic weights that we calculated in Section~\ref{sec:mitigation}.}

\subsubsection{Comparison with SDSS DR16}
\begin{figure}
   \includegraphics[scale=1]{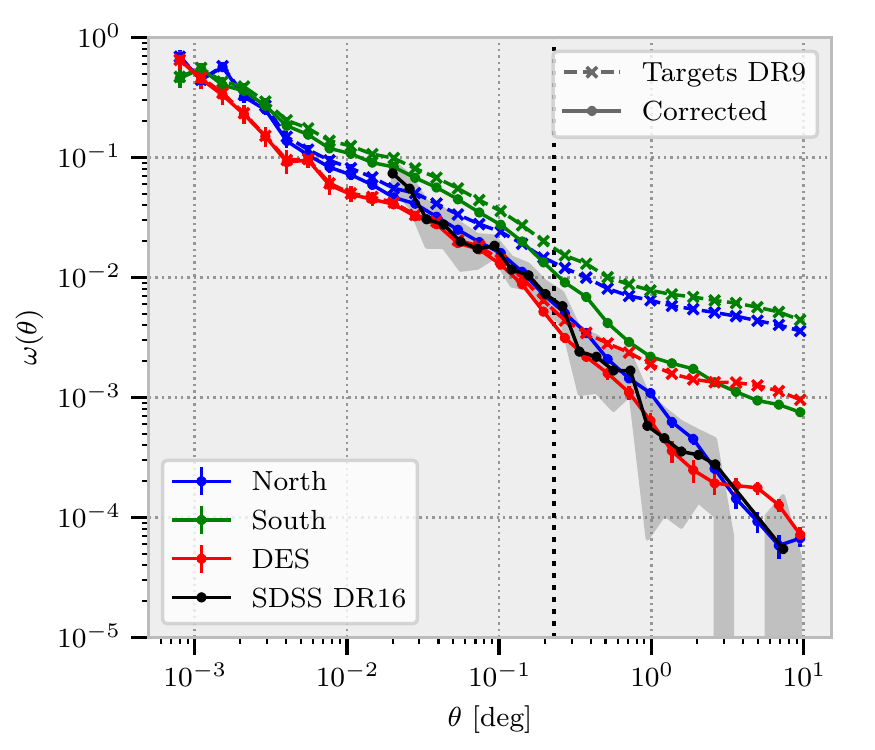}
    \centering
    \caption{The angular 2-point correlation function of the DR9 QSO targets.  The dashed lines represent the nominal DESI QSO target selection. The solid lines are for the corrected QSO target selection with RF method.  The black line is the angular correlation function from SDSS DR16 (which cannot be calculated at angles smaller than $62~\rm{arcsec}$, due to fiber collisions).  The grey region despicts the error for SDSS DR16 estimated as the standard deviation across 100 EZ-mocks.  The black dotted line corresponds to the typical resolution of our regression analyses \ie the pixel size at $N_{\rm side}=256$. The solid green line is discussed in detail in Appendix\,\ref{sec:SgrStream}.}
    \label{fig:ang_corr_dr9}
\end{figure}

We calculate the angular correlation function of the raw and corrected (RF method) QSO target densities for the three different imaging footprints.  For comparison,  we also calculate the angular correlation function of  SDSS DR16 quasars.  Note that the correlation with SDSS quasars cannot be computed at angles smaller than 62 arsec due to fiber collisions \citep[e.g.][]{Dawson2016}. For SDSS DR16 quasars, we use the systematic weights provided in \cite{Rezaie2021} based on a neural network treatment \citep[see][for the standard treatment]{Ross2020}. The results are shown in Figure\,\ref{fig:ang_corr_dr9}.  Note that the error bars shown in Figure\,\ref{fig:ang_corr_dr9} are the standard deviation of the LS estimator \citep{Landy1993} except for the correlation function of SDSS quasars, for which the errors are estimated using the standard deviation across 100 EZ-mocks from \cite{Zhao2021}. 

After mitigating for systematics, the angular correlation in the North and in the DES region (again, see Figure\,\ref{fig:psfdepth_r} for the definition of these regions) are comparable with the correlation computed with the SDSS DR16 sample. The SDSS DR16 sample has been carefully corrected for systematic effects (as derived in previous work), and so is expected to be largely free from any contamination. In the South,  even after mitigating for systematics, we do not recover the same level of correlation as for the SDSS DR16 sample.  Reasons for this difference are discussed further in Appendix\,\ref{sec:SgrStream}.

It is worth noting that the correction is larger at larger angles, \ie the impact of systematic effects is higher on large scales. Therefore,  mitigation of photometric systematics is critical for studies that require information from large scales, such as studies of primordial non-Gaussianity.

\subsubsection{Regression method comparison}
\label{sec:method_comparaison}
\begin{figure}
   \includegraphics[scale=1]{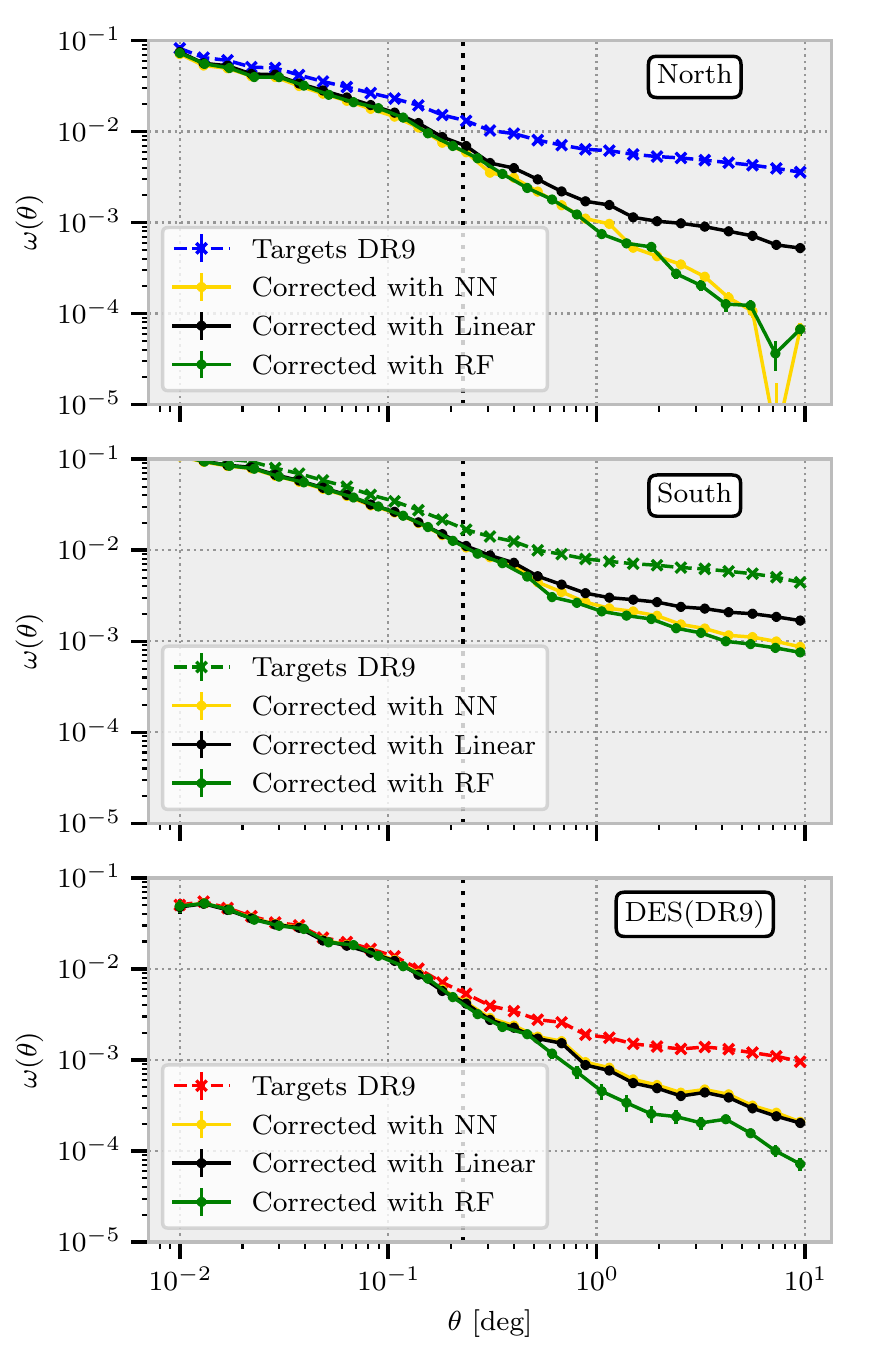}
    \centering
    \caption{The angular correlation function for the three different methods introduced in Section~\ref{sec:mitigation} for the three regions highlighted in Figure\,\ref{fig:psfdepth_r}.  The Linear method is less \rc{effective} than the two machine-learning-based methods, especially in the most strongly contaminated region (the North).  The correction in the South \rc{does not improve properly the correlation } for any of the three methods.  The RF and NN methods produce similar results in the North but  slightly different results in DES. This is because the RF is more robust to sampling high levels of contamination from a small number of pixels.}
    \label{fig:ang_corr_method_comparaison}
\end{figure}

Figure~\ref{fig:ang_corr_method_comparaison} shows the difference between the three methods introduced in Section~\ref{sec:mitigation} on the North,  on the South and on DES.  As expected,  the linear method is less \rc{effective} than the other two in the highly contaminated North (top panel), highlighting the necessity of our more complex machine-learning-based regressions. The two machine-learning-based methods give similar results in the North.  None of the three methods properly correct for the contamination in the South (middle panel); this is discussed further in Appendix\,\ref{sec:SgrStream}.

The information in Figure \ref{fig:systematics_plot_des} suggests that all three methods introduced in Section~\ref{sec:mitigation} are quite \rc{effective} at mitigating systematics in the DES region. However, the bottom panel of Figure~\ref{fig:ang_corr_method_comparaison} demonstrates that the linear method and the NN method less \rc{effectively} correct the angular correlation function in the DES region as compared to the RF method. The difference between the linear and the RF methods comes from the non-linear part of the contamination in this region.  The systematics plots show that in the highly contaminated regions, the linear method \rc{corrects less than the RF.}

The difference between the NN and the RF is more subtle since the training information, \ie the area and the chosen folds in that area,  is  similar.  The explanation comes from the difference between the two algorithms. The NN method is less efficient to correct small, highly contaminated regions. The RF creates boxes in  feature space and separates the most contaminated regions from the rest. The estimation of the correction weight is then possible everywhere. 

To solve this problem, a regularization term can be introduced to force the NN to also consider  small regions. The choice of the regularization value strongly depends on the size of the small regions  and on their location  in  feature space.  Without realistic mocks for the DESI QSO sample,  this additional hyper-parameter cannot be easily optimized. The training time for the NN also varies considerably as a function of the value of the tolerance chosen to sample the NN hyper-parameters, whereas, with our parameterisation,  the RF method is quicker to train.
\rc{Since the RF correction already obtains good results in DES,  we do not need to improve the current NN method.  We leave fine optimizing of the NN method for future work when realistic DESI QSO mocks become available.}

\subsubsection{Resolution of the correction}
\label{sec:256vs512}
The size of the pixels used when determining weights to mitigate systematics is critical, because it gives the scale at which the correction is most \rc{effective}.  We perform all the analysis above with $N_{\rm side}=256$ corresponding to a characteristic angle of $\sim$0.22 deg \ie 12.6 $h^{-1}$Mpc at $z \sim 1.7$ \citep[Planck 2015;][]{Ade2016}.  This pixel size is chosen such that there are sufficient targets inside each pixel to learn about contamination without introducing significant Poisson noise in the mitigation. 

For the DESI QSO target selection,  a pixel with a size of $N_{\rm side}=256$ typically contains a median of $\sim$16 QSO targets.  At a size of $N_{\rm side} = 512$,  the number of targets decreases to $\sim$4. With such a small number of targets,  the per-pixel density is too noisy for  machine learning methods to relate fluctuations in density to observational features. For the same reason, we needed to increase the pixel-size used in our analysis to $N_{\rm side} = 128$ when studying eBOSS mocks, since the density of the eBOSS mocks was lower than the DESI QSO target selection density (see Appendix~\ref{sec:mocks}).  

Other DESI targets, such as ELGs and LRGs, will have a higher density and it will be possible to decrease the size of the pixels used to estimate correction weights by a factor of 2--4.

\subsubsection{Systematic checks: restrictive QSO target selection}
\label{sec:restrictive}
Since the DESI QSO targets are selected with an RF classification, the stellar contamination depends on the value of the selection threshold. This dependence propagates into the angular correlation,  since we broadly expect a lower $r_0$ for a less contaminated sample.  

\begin{figure}
    \includegraphics[scale=1]{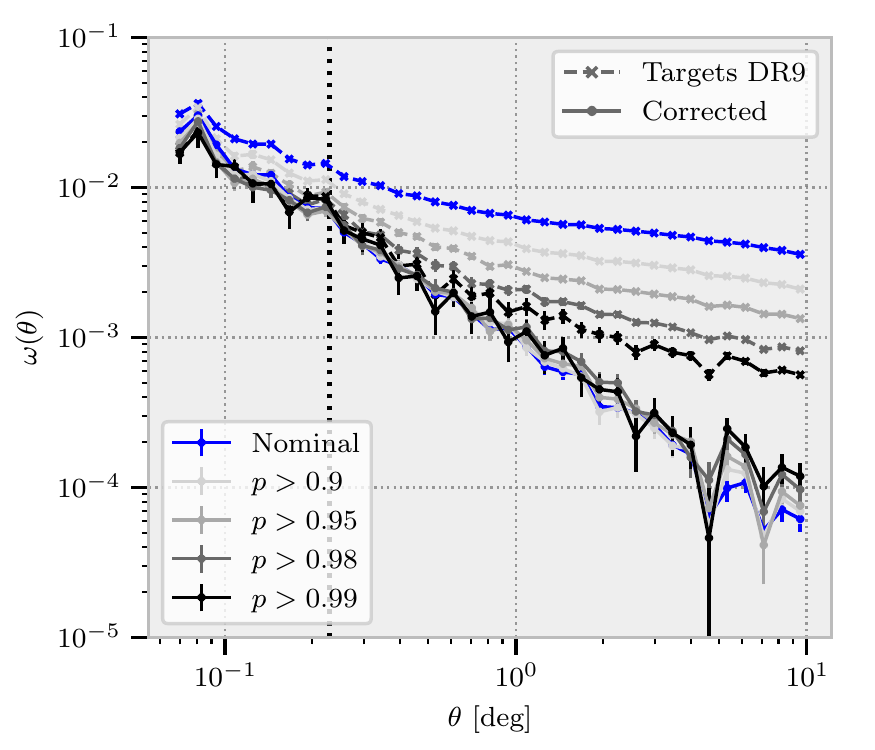}
    \centering
    \caption{Evolution of the angular correlation function in the North region when the probability threshold of the quasar target selection is increased.  \rc{The nominal probability in the North is $p(r)= 0.88 - 0.04 \times \tanh(r - 20.5)$}. The dashed lines depict raw target selections with different probability thresholds. The solid lines show the corrected versions of the same selections.  The black dotted line corresponds to the typical resolution of the correction \ie the pixel size at $N_{\rm side}=256$.}
    \label{fig:ang_corr_restrictive}
\end{figure}

Figure~\ref{fig:ang_corr_restrictive} shows the angular correlations in the North for DESI QSO targets with different values of the selection threshold (dashed lines) and of the corresponding corrected samples (full lines).  \rc{The systematic weights are generated with the RF method for each target selection. } The nominal probability in the North footprint used in the QSO selection depends on $r$ and is given by $p(r) = 0.88 - 0.04 \times \tanh(r - 20.5)$.  This threshold is lower than for the other regions.  When the threshold increases,  the selection is less contaminated since the RF will select a higher fraction of bonafide QSOs.

The amplitude of the angular correlation in Figure~\ref{fig:ang_corr_restrictive} decreases as the probability threshold is increased, suggesting that the excess of correlation is mostly due to stellar contamination. This also confirms  that our method to mitigate the systematics does not over-fit the data since when the contamination is removed,  the correlation converges to the same level as the SDSS QSO correlation (cf. Figure~\ref{fig:ang_corr_dr9}).

Finally, we note that the excess correlation in the South is {\em not} reduced when the probability threshold of the RF selection is increased. This is because many stars in the Sagittarius Stream have selection features that resemble QSOs. These stars are assigned near unit probability by the DESI QSO targeting algorithm, such that they are not removed by a more restrictive selection.

\subsection{Limber parameters}
\subsubsection{Limber approximation}
\label{sec:limber}
The angular correlation function $w(\theta)$ is related to the 3D one $\xi(r)$ through:
\begin{equation}
w(\theta) = \int_0^\infty \int_0^\infty \xi(r_{12}) \tilde{S}(r_1)r_1^2 dr_1 \tilde{S}(r_2) r_2^2  dr_2, 
\end{equation}
where $r_{12} = \sqrt{r_1^2 + r_2^2 - 2 r_1 r_2 \cos(\theta)}$ and $\tilde{S}(r)$  represents the probability for an object at distance $r$ to be observed by the survey. It is defined from the selection function $S(r)$ by 
\begin{equation}
\tilde{S}(r) = \dfrac{S(r)}{\int_0^{\infty} S(r) r^2 dr}. 
\end{equation}

\noindent We will estimate $S(r)$ by multiplying the Quasar Luminosity Function from \cite{Palanque-Delabrouille2013} by the completeness of the DESI QSO target selection.

Adopting the approximation that the correlation function is non-negligible only for small values of $r$ \ie that the angular correlation function is non zero only for small values of $\theta$, one can assume that $x = r_2 - r_1$ and $y = (r_1 + r_2) / 2$,  and hence derive Limber's equation valid only for small angles \citep[for an explicit derivation,  see for instance][]{Kurki-Suonio2019}: 
\begin{equation}
w(\theta) \simeq \int_0^\infty \tilde{S}(y)^2 y^4 \int_{-\infty}^{\infty} \xi\left(\sqrt{x^2 + y^2 \theta^2} \right) dx dy. 
\end{equation} 

Assuming a power-law for the correlation function: 
\begin{equation}
\xi(r) = \left( \dfrac{r}{r_0}\right)^{- \gamma},  
\end{equation}
the angular correlation function becomes:
\begin{equation}
w(\theta) = \sqrt{\pi}  \dfrac{\Gamma\left(\frac{\gamma - 1}{2}\right)}{\Gamma\left(\frac{\gamma}{2}\right)} r_0^{\gamma} \theta^{1 - \gamma}  \int_0^{\infty} \tilde{S}^{2}(y) y^{5 - \gamma} dy
\end{equation}
\noindent where $\Gamma$ is the gamma function.

The integral calculation requires a fiducial cosmology.  \rc{We choose a $\Lambda$ cold dark matter ($\Lambda$CDM) cosmology following the Planck 2015 parameters from \citet{Ade2016}: $\Omega_{m,0}=0.308$,  $\Omega_{\Lambda, 0}=0.691$,  $\Omega_{b, 0}=0.048$,  $h=0.677$,  $\sigma_8=0.815$,  $n_s = 0.967$. }

\subsubsection{Fitting of Limber parameters}
As explained in Section~\ref{sec:limber}, the power-law parameterisation of the correlation function $\xi(r)$ can be constrained by the angular correlation function $w(\theta)$ at small angles. We proceed by estimating these power-law parameters and comparing them to previous measurements done with SDSS data by \cite{Myers2009} and with 2dF data by \cite{Croom2005}.

\begin{table*}
\centering
\caption{Limber parameters for the DESI QSO target selection in the imaging footprints depicted in Figure\,\ref{fig:psfdepth_r}.  The errors are estimated using a sub-sampling method.  We provide the measurements for both the non-corrected and corrected cases. \rc{The correction is performed with the RF method. } $\bar{z}$ is the mean redshift  of the sample. }
\label{tab:limber}
\begin{tabular}{lcccccllll}
 &
  $r_p$ min $[\rm{h}^{-1} \rm{Mpc}]$ &
  $r_p$ max $[\rm{h}^{-1} \rm{Mpc}]$ &
  $\bar{z}$ &
  $r_0$ $[\rm{h}^{-1} \rm{Mpc}]$ &
  $\gamma$ &
  $r_0$ $[\rm{h}^{-1} \rm{Mpc}]$ &
  $\gamma$ &
   &
   \\[1mm] \cline{1-8} \\[-2mm]
Croom &
  1.0 &
  25 &
  1.35 &
  \multicolumn{1}{c}{-} &
  \multicolumn{1}{c}{-} &
  $5.84 \pm 0.33$ &
  $1.64 \pm 0.04$ &
   &
   \\
Myers &
  1.6 &
  40 &
  2 &
  \multicolumn{1}{c}{-} &
  \multicolumn{1}{c}{-} &
  $4.56 \pm 0.48$ &
  $1.5$ (fixed) &
   &
   \\[1mm] \cline{1-8} \\[-2mm]
 &
  \multicolumn{1}{l}{} &
  \multicolumn{1}{l}{} &
   &
  \multicolumn{2}{c}{DESI QSO Targets} &
  \multicolumn{2}{c}{Corrected} &
  \multicolumn{1}{c}{} &
  \\[1mm] \cline{1-8} \\[-2mm]
North &
  0.045 &
  45 &
  1.7 &
  $10.15  \pm 0.70$ &
  $1.61  \pm 0.03$ &
  $7.49  \pm 0.57$ &
  $1.89  \pm 0.02$ &
   &
   \\
South &
  0.045 &
  45 &
  1.7 &
  $12.88  \pm 0.95$ &
  $1.64 \pm 0.03$ &
  $10.33  \pm 0.84$ &
  $1.80 \pm 0.02$ &
   &
   \\
DES ($\dec > - 30$) &
  0.045 &
  45 &
  1.7 &
  $6.76 \pm  0.58$ &
  $1.78  \pm 0.04$ &
  $6.15  \pm 0.45$ &
  $1.88 \pm 0.04$ &
   &
   \\
DES &
  0.045 &
  45 &
 1.7 &
  $7.19  \pm 0.34 $ &
  $1.79   \pm 0.03 $&
  $6.47  \pm 0.31 $&
  $1.89  \pm 0.02 $&
   &
\end{tabular}
\end{table*}

\begin{figure}
    \includegraphics[scale=1]{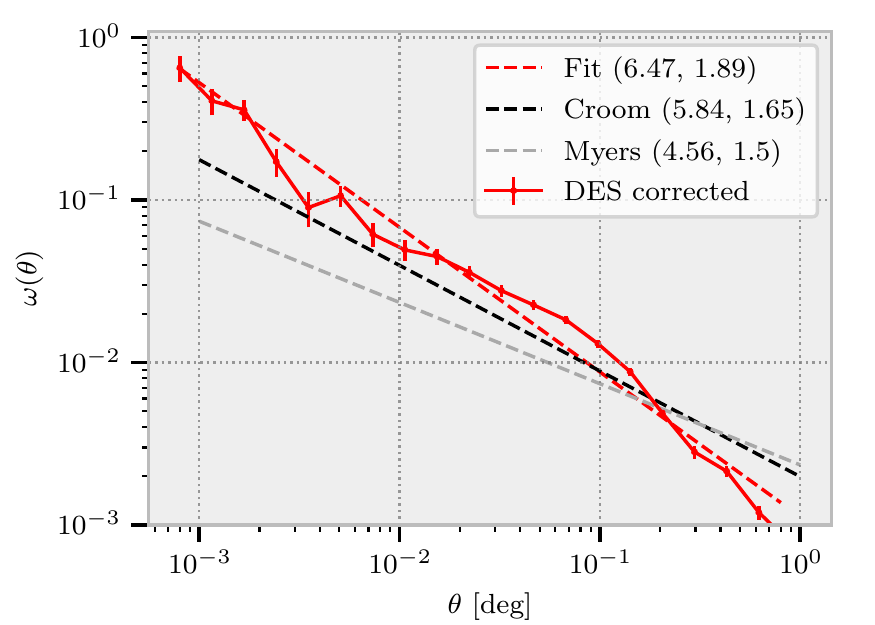}
    \centering
    \caption[Limber parameters in DES.]{Fit of the Limber parameters in the DES region for the corrected DESI QSO targets (red line).  The Limber parameters $(r_0, \gamma)$ are given in each case.  The black (\textit{resp.} grey) dashed line shows the Limber function measured by \cite{Croom2005} \citep[\textit{resp.}][]{Myers2009}. The parameters for the other regions are given in Table~\ref{tab:limber}. }
    \label{fig:limber}
\end{figure}

The Limber parameters are estimated in the three imaging regions highlighted in Figure~\ref{fig:psfdepth_r} and \rc{also in the DES region with only $\dec > -30$,  which we will refer to as DES($\dec > -30$).  This region will almost correspond to the intersection between DES and the expected nominal DESI footprint.} The results are shown in Table\,\ref{tab:limber} and fit for the corrected DESI QSO targets in DES region is plotted in Figure~\ref{fig:limber}.  The correlation function is fitted from $1e^{-3}$ to $0.8\,\rm{deg}$,  corresponding to a transverse separation of $0.045$ to $45~\rm{h}^{-1}\rm{Mpc}$ at redshift $1.7$.  \rc{The mean redshift of the DESI QSO targets ($\sim 1.7$) is obtained tanks to the selection function introduced in Section \ref{sec:limber} which corresponds to the estimated redshift distribution of the sample.}

The errors on the parameters $r_0$ and $\gamma$ are estimated using a sub-sampling method: 10 (\textit{resp.} 18, 10, 4) sub-regions of similar area ($\sim$ 450 deg$^2$) are used for the North (\textit{resp.} South,  DES,  DES $\dec> - 30$)). Sub-sampling is used to probe the variability of the angular correlation function in different areas of the footprint, where stellar contamination and systematic effects should differ.  \rc{If the systematics are properly mitigated,  the error of each parameter should be reduced \ie the corrected angular correlation functions over each sub-region should be more similar. }

The value $r_0$, which parameterises the amplitude of the correlation function, captures the offset of different measured angular correlation functions. For instance, the fact that the angular correlation in the South is higher than in DES, manifests in a value of $r_0$ that is higher in the South than in DES.  As systematics tend to lead to an increase in amplitude, $r_0$ tends to be higher when systematics are dominant. For example, the fact that the value of $r_0$ that we measure in the DES footprint is generally comparable to the values found for  previous measurements \citep{Croom2005,  Myers2009}, suggests that systematics have been mitigated well in the DES region.

The parameter $\gamma$ describes the slope of the angular correlation.  A higher value of $\gamma$ means a steeper slope.  It is worth noting that since the  \rc{relevant correction scales of the systematic weights} are larger than the pixel size used for our analyses ($\sim0.2\,\rm{deg}$),  the angular correlations of the corrected targets have a steeper slope than for the raw target samples.  The comparison of $\gamma$ between the DESI QSO targets and previous measurements is  not particularly relevant since any excess correlation caused by systematics on small scales, \ie below the typical angular resolution of our corrections, cannot be removed with our  mitigation procedure.

Though no spectroscopic confirmation of DESI QSO targets is available across a significant fraction of the footprint used in our study,  our correction for angular systematics enables us to provide \rc{a Limber amplitude parameter $r_0$ which is consistent with those found by \citet{Croom2005, Myers2009} using spectroscopic data. } Nevertheless, the fitting of the Limber parameters will be greatly improved by information from the DESI spectroscopic survey. Follow-up spectroscopy will enable us to remove stars from the DESI QSO target sample, so any clustering analyses will be free of the stellar contamination that increases the amplitude of the angular correlation measurements. 

\section{Conclusion}
\label{sec:conclusion}
The QSO target selection for the five-year DESI survey, which is based on the ninth Data Release (DR9) of the Legacy Imaging Surveys, has recently been finalized.
This study \rc{illustrates} imaging systematic effects and stellar contamination in the DESI QSO selection, with a view to studying and correcting for large-scale density fluctuations that will impact DESI clustering measurements.

After a brief description of the current QSO target selection for DESI,  we explain the different effects that cause density variations in the target sample. These effects can be separated into three categories:
\begin{itemize}
\item[$\blacktriangleright$] Stellar contamination from the Milky Way. This occurs mainly near to the Galactic plane and is caused by the large excess of stars in this region. 
\vspace{3mm}
\item[$\blacktriangleright$] Impact of the depth of the observations in the $z$ and $W2$ bands. These two bands are crucial for DESI QSO target selection.  Hence,   $z$-band systematics and above-average $W2$ PSF Depth produce strong over-densities, which are most pronounced in the North of the Legacy Surveys imaging.  \rc{The impact of the depth in the $W2$ band is hidden in the South  by the numerous stars from the Milky Way.}
\vspace{3mm}
\item[$\blacktriangleright$] Stellar contamination from the Sagittarius Stream.  These stars have similar colours to QSOs and thus heavily contaminate the DESI QSO target selection. A map of the Sagittarius Stream enables us to remove the contamination to first-order but is insufficient to correct for higher-order effects, as demonstrated by the angular correlation function of DESI QSO targets in this region. \rc{The part of the remaining correlation which comes from Sagittarius Stream stars cannot currently} be removed with imaging maps alone,  neither by improving the QSO target selection method nor by improving the Sagittarius Stream map.  This is because many stars in the Stream are too faint to have been previously observed.  \rc{These stars prevent us from properly analyzing the impact of other imaging features on target selection and the excess correlation observed in this region.  However, this analysis will be done once these stars are removed by spectroscopy with DESI.}
\end{itemize}

The QSO target selection, incorporating weights to correct for systematics, was validated by computing the angular correlation function in each imaging region of the DESI Legacy Imaging Surveys.  We show that after mitigating for systematics,  the angular correlation function from the target selection  is comparable to the angular correlation function of SDSS DR16 quasars, except for in the Sagittarius Stream region (as explained above). Our results are very encouraging for the upcoming DESI spectroscopic survey, since our analysis can produce similar angular correlation functions as derived for SDSS spectroscopic data, despite utilizing only information from imaging.

The method described in this paper was built and optimized specifically for the DESI QSO target selection.  In particular, the hyper-parameters for the NN and the RF methods are specific to DESI QSO targets. Nevertheless, our approach could readily be adapted to any DESI target class. In particular,  the form of the K-fold training  was developed independently of the target class.  

Once the DESI survey progresses,  we will apply the method outlined in this paper to compute imaging systematic correction weights for the DESI spectroscopic QSO catalogue.  We expect these corrections to be smaller for future DESI samples, since many first-order  systematic effects that are caused by stellar contamination will be removed by spectral information. 

\section*{acknowledgements}

This research is supported by the Director, Office of Science, Office of High Energy Physics of the U.S. Department of Energy under Contract No. DE–AC02–05CH11231, and by the National Energy Research Scientific Computing Center, a DOE Office of Science User Facility under the same contract; additional support for DESI is provided by the U.S. National Science Foundation, Division of Astronomical Sciences under Contract No. AST-0950945 to the NSF’s National Optical-Infrared Astronomy Research Laboratory; the Science and Technologies Facilities Council of the United Kingdom; the Gordon and Betty Moore Foundation; the Heising-Simons Foundation; the French Alternative Energies and Atomic Energy Commission (CEA); the National Council of Science and Technology of Mexico; the Ministry of Economy of Spain, and by the DESI Member Institutions.

ADM was supported by the U.S. Department of Energy, Office of Science, Office of High Energy Physics, under Award Number DE-SC0019022.

The authors are honored to be permitted to conduct astronomical research on Iolkam Du’ag (Kitt Peak), a mountain with particular significance to the Tohono O’odham Nation.

\section*{Data availability}
The Data Release 9 of the DESI Legacy Imaging Surveys is available via \url{https://www.legacysurvey.org/dr9/}. 
The systematic weights derived in this article will be shared on reasonable request to the corresponding author.


\bibliographystyle{mnras}
\interlinepenalty=10000
\bibliography{content/bibli}

\appendix
\interlinepenalty=0
\section{Validation with mocks}
\label{sec:mocks}
To validate the systematic mitigation method and avoid overfitting,  we use a set of 100 QSO EZ-mocks from eBOSS \citep{Zhao2021}.  They have a smaller density than the nominal target selection density and a smaller area.  However,  it will not impact the analysis since the surface and the density are enough for our test.  Besides,  since our analysis for the legacy survey is different for the three photometric footprints,  \rc{we will mimic one footprint only,  similar to a sub-region of the North footprint.}

The mocks contain a cosmological signal but no systematic effects. They are contaminated using the inverse weight estimated previously and this contamination will be mitigated with our method using the same observational features.

We present here contamination estimated with the RF method (it is the inverse of the map shown in Figure~\ref{fig:weight}).  \rc{We test the method in two different ranges of contamination: one describing a weakly contaminated case (extracted from the  $\ra,\dec \in [100^\circ, 270^\circ] \times [32^\circ,  60^\circ]$ box) where all the methods work well; a second describing a strongly contaminated case (extracted from the $\ra,\dec \in [120^\circ, 290^\circ] \times [55^\circ,  90^\circ]$ box) where the linear method} is inefficient as shown in the systematics plots of Figure~\ref{fig:systematics_plot_north}.  To compare the method efficiency and avoid biasing the results,  we also applied an NN-based estimation for the strong contaminated case.

\begin{figure}
    \includegraphics[scale=0.4]{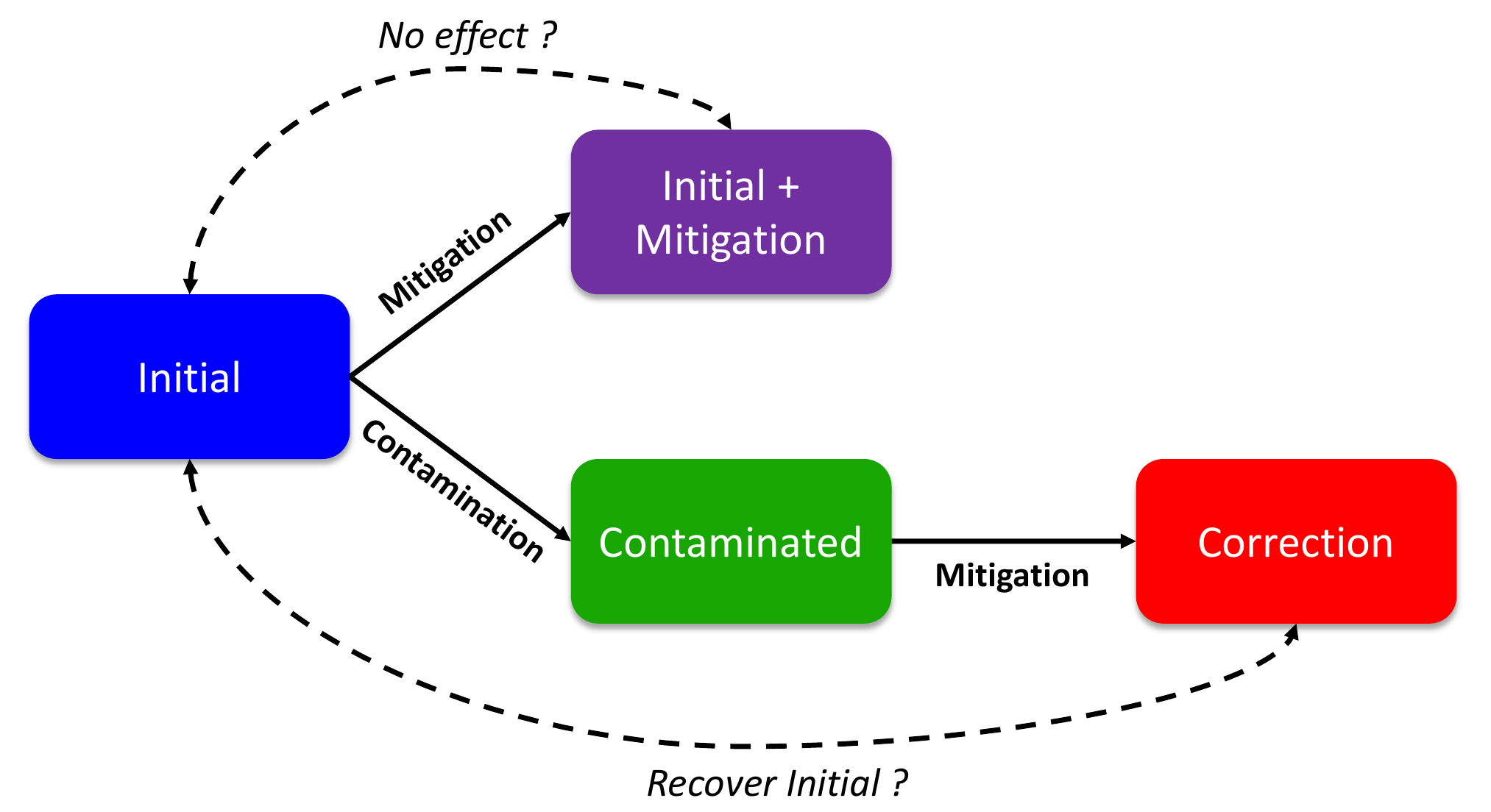}
    \centering
    \caption{Pipeline validation with mocks.  The initial mocks (blue) are uncontaminated and contain a cosmological signal. They are contaminated with systematic weights (green). Then, they are corrected with the mitigation presented above (red).  We compare the corrected mock to the initial mock to verify if we recover the correct initial state.  As a sanity check,  we also applied the mitigation to the initial mock (purple) to verify whether our mitigation technique overfits target density variations. These tests are represented by the two black dashed lines.}
    \label{fig:pipeline}
\end{figure}

The validation pipeline is shown Figure~\ref{fig:pipeline}. There are two different tests (dashed lines). The first one is to check whether our method overfits the data.  We apply the mitigation method to the initial (uncontaminated) mocks such that no correction is expected. The angular correlations of the uncontaminated mocks and of the corrected uncontaminated mocks should be identical. The second test is to validate our mitigation. We apply the mitigation method to the contaminated mocks, expecting to recover the same angular correlation as for the initial mocks.

Some subtleties have to be considered.  First, the density of these mocks ($70~\rm{deg}^{-2}$) is lower than that of the quasar targets ($300~\rm{deg}^{-2}$).  Thus, the size of the correction cannot be the same since the number of objects inside a pixel of $N_{side} = 256$ is too small. We therefore downgrade the contamination and the observational feature maps to $N_{side} = 128$.

Secondly,  the footprint of these mocks is different from the North footprint.  The size of the training sample is here twice smaller as the data.  We, therefore, use only 3 folds instead of 6 for the DR9 North training.  It is worth noting that we need to change the position of the folds to cover the footprint correctly as explained in \ref{sec:k-fold}, especially for the strongly contaminated case. 

Nevertheless,  the training conditions are slightly different and we expect some differences for the second test.   

\begin{figure}
    \includegraphics[scale=1]{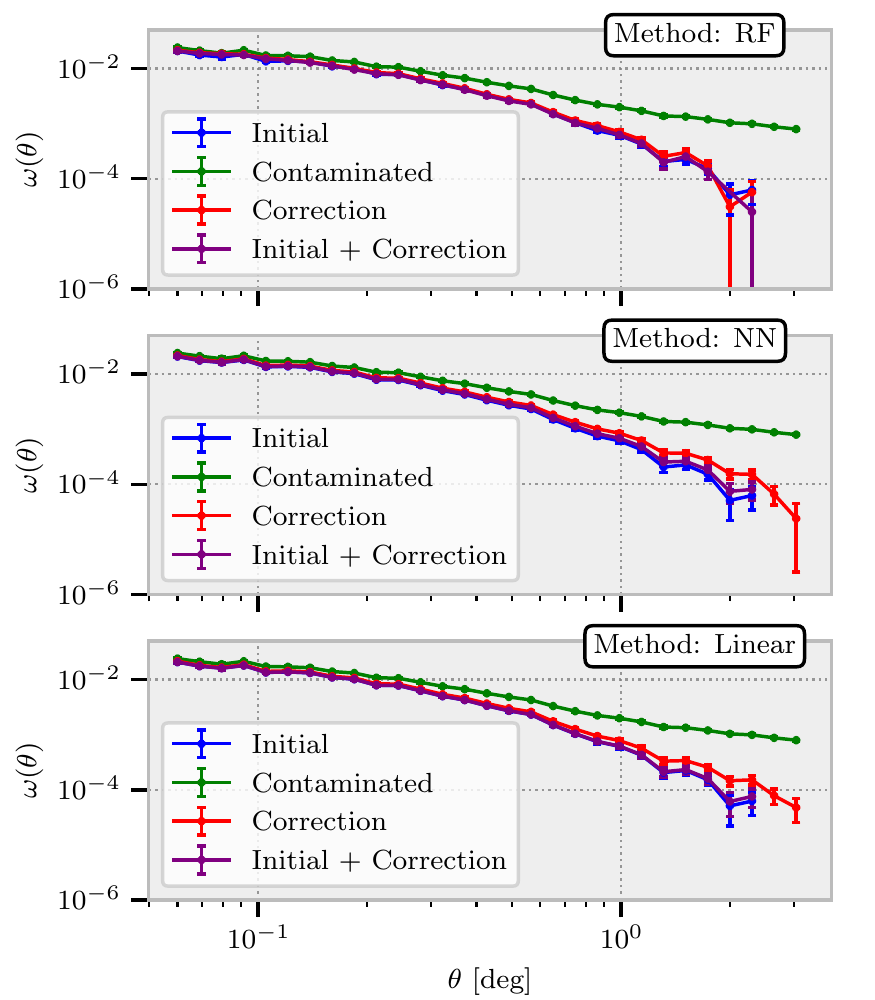}
    \centering
    \caption{\rc{Weakly contaminated case: Mean on 100 eBOSS EZ-mocks of the angular correlation functions.} The error bars are the standard deviation between all realizations. Colours follow the same scheme as in Figure~\ref{fig:pipeline}. The systematic contamination is done with a RF estimation of a weakly contaminated area ($\ra,\dec \in [100, 270] \times [32,  60]$) of the DR9 legacy survey.  The method does not overfit the data and recovers the initial correlation.  Note that the contamination was estimated with the RF method which explains the small residual systematics with the NN and Linear correction. }
    \label{fig:mocks_v1}
\end{figure}

The angular correlation functions for the weakly contaminated case are shown in the Figure~\ref{fig:mocks_v1}.  Each correlation function is the average of the angular correlations of the 100 mocks and the error bars are the standard deviation between all the realizations.  We show the correlation function up to $\theta \sim 3.5 ~\rm{deg}$,  which corresponds to a comoving angular distance of $\sim 250~\rm{Mpc}~\rm{h}^{-1}$ at $z=1.7$,  the maximum size of the eBOSS analysis.  The result for each regression method is shown in the three panels. 

The purple lines are the angular correlations of the initial mocks mitigated with our methods.  In the three cases,  they lie on top of the blue lines that show the angular correlations of the initial mocks. \rc{This indicates no sign of overfitting compare to the significance of the correction}.  The green lines are the angular correlations for the mocks after contamination. The red lines are the angular correlations for the contaminated mocks after mitigation.  The red line exactly recovers the blue line in the case of the RF method.  In the two other cases,  the red lines recover more or less the blue lines and the slight difference stems from the contamination was done from an estimation of the systematics by the RF method which benefits the RF case.

\begin{figure}
    \includegraphics[scale=1]{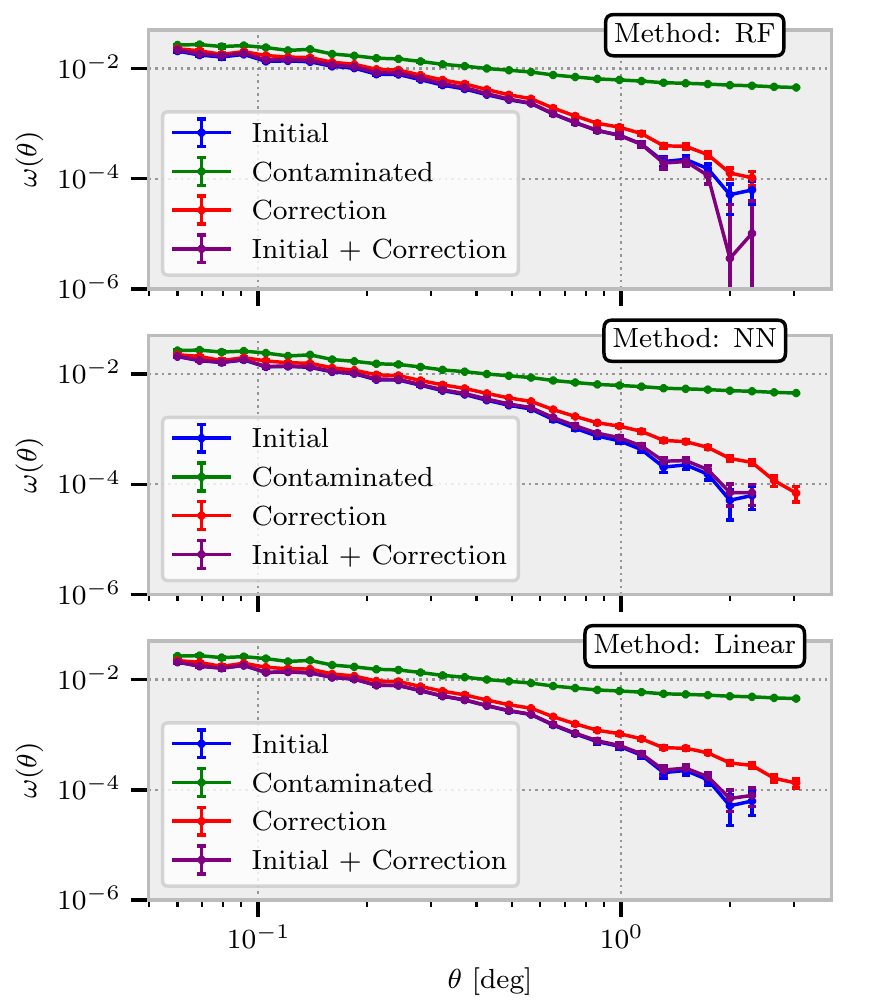}
    \centering
    \caption{Strongly contaminated case: Same as Figure~\ref{fig:mocks_v1} but the systematic contamination depicts strong contaminated area ($\ra,\dec \in [120, 290] \times [55,  90]$). The correction is not perfect since the training sample is smaller than the Legacy Surveys case and cannot contain the same information in each fold.  Besides,  the correction is done with a higher $N_{side}$ which can explain that we do not fully recover the initial state. }
    \label{fig:mocks_v2}
\end{figure}

\begin{figure}
    \includegraphics[scale=1]{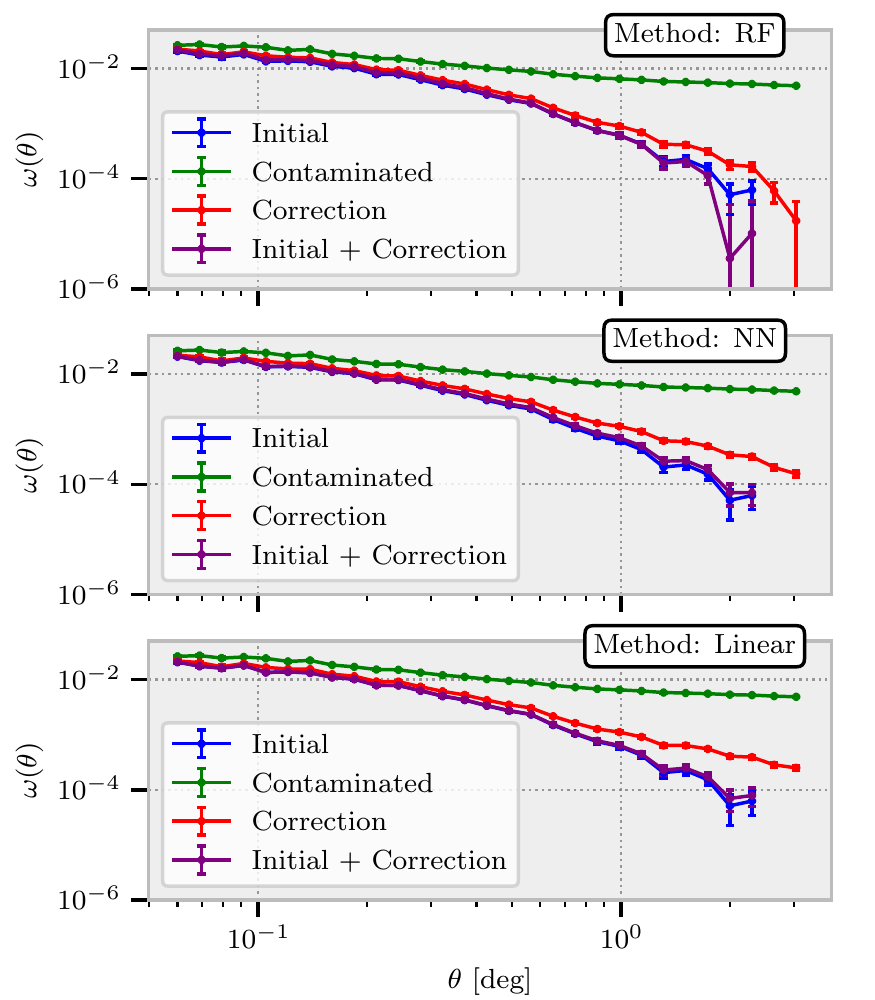}
    \centering
    \caption{Strongly contaminated case: Same as Figure~\ref{fig:mocks_v2} but the systematic contamination which depicts strong contaminated area ($\ra,\dec \in [120, 290] \times [55,  90]$) is estimated with the NN method. }
    \label{fig:mocks_v2_with_nn}
\end{figure}

The angular correlations for the strongly contaminated case are shown in Figure~\ref{fig:mocks_v2} for the RF-based contamination and in Figure~\ref{fig:mocks_v2_with_nn} for the NN-based one. 
\rc{The two figures are similar,  indicating that the contaminated method does not benefit the RF or NN mitigation.  The result from the same contamination method can be compared together.}

\rc{Here again, there is no indication of overfitting for the NN and Linear method.  For the RF method,  the purple line is slightly below the blue one above $1 \deg$ indicating tiny overfitting.  Note that the mocks that we used here cover a smaller area than the DESI QSO sample,  thereby making our method more prone to overfit for this test case.  Optimization of the hyper-parameters to reduce overfitting on the DESI footprint will be pursued with DESI mocks.  However,  this tiny overfitting is much less significant than the level of systematic corrections validating that most of the impact of RF weights on the angular correlation function of the DESI QSO sample is real systematics mitigation and not overfitting. }
 
 The angular correlation of contaminated mocks is higher as a consequence of the stronger contamination compared to the previous case. The red lines do not perfectly recover the blue lines even in the case of the RF regression.  Indeed,  the smaller size of the training sample and the modification of the fold form prevent the NN and RF methods to learn all the information needed to fully correct this contamination. In addition,  the NN method is less efficient to correct small contaminated regions without any additional hyper-parametrization like regularization terms.  The RF approach is more robust than the NN one to variations in the training sample and less dependent on hyper-parametrization.  For the linear method,  the correction is less efficient: in the region of extreme observational features,  the systematic effects are less corrected,  as shown in Sec. ~\ref{sec:systematic_plots}.

This set of mocks and the second test are used to optimize the hyper-parameters of the NN and perform the grid search method (\ref{sec:NN}).  Given the subtleties presented above, however, it is expected that the hyper-parameters used here are not the best one for the DR9 training.  More optimized parameter tuning will eventually be achieved with mocks matching the DESI QSO samples.

\section{South footprint and Sagittarius Stream contamination}
\label{sec:SgrStream}

\begin{figure}
    \includegraphics[scale=1]{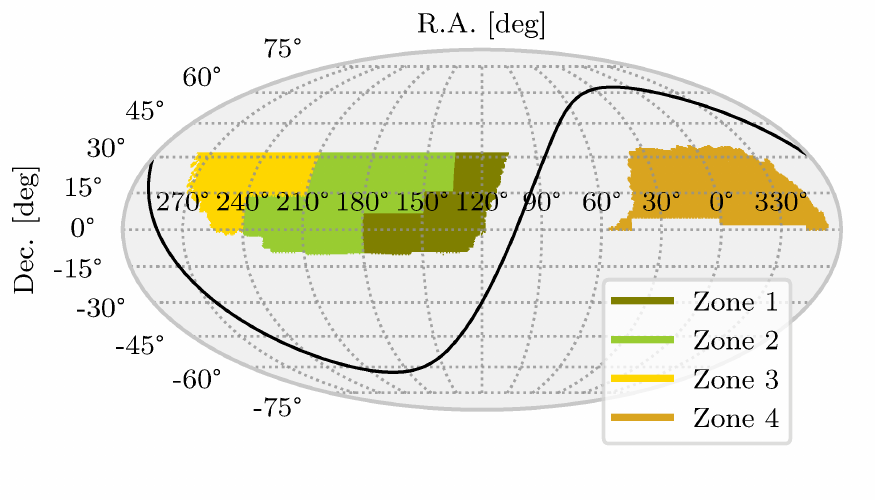}
    \centering
    \caption{The South region is split in four distinct zones to analyse the stars contamination.  Zone 1 near the anti-galactic pole shows less over density than the three others.  Zone 2 contains the Sagittarius Stream.  Zone 3 and Zone 4 shows strong over density due to stars from the galactic plane.}
    \label{fig:zone_repartition}
\end{figure}

As shown in Figure~\ref{fig:ang_corr_dr9},  even after the systematic mitigation,  we do not recover the same level of correlation in the South footprint as in the two other regions.  To analyze the excess of correlation in the South footprint,  we divide it into 4 zones,  represented in Figure~\ref{fig:zone_repartition}.  These zones are:
\begin{itemize}
\item[$\blacktriangleright$] Zone 1 near the anti-galactic pole shows less over-density than the other regions near the Galactic plane.  Its lower density is due to the lower value of the PSF Depth $W2$ in this zone.
\vspace{3mm}
\item[$\blacktriangleright$] Zone 2 contains the Sagittarius Stream and it is strongly contaminated by stars.
\vspace{3mm}
\item[$\blacktriangleright$] Zone 3 and zone 4 show strong over-density due to stars from the galactic plane.  In addition,  zone 4 describes the SGC part of the South footprint and we considered also zone 13 combining zone 1 with zone 3 which describes the NGC without the Sagittarius Stream part of the South.
\end{itemize}

\begin{figure}
    \includegraphics[scale=1]{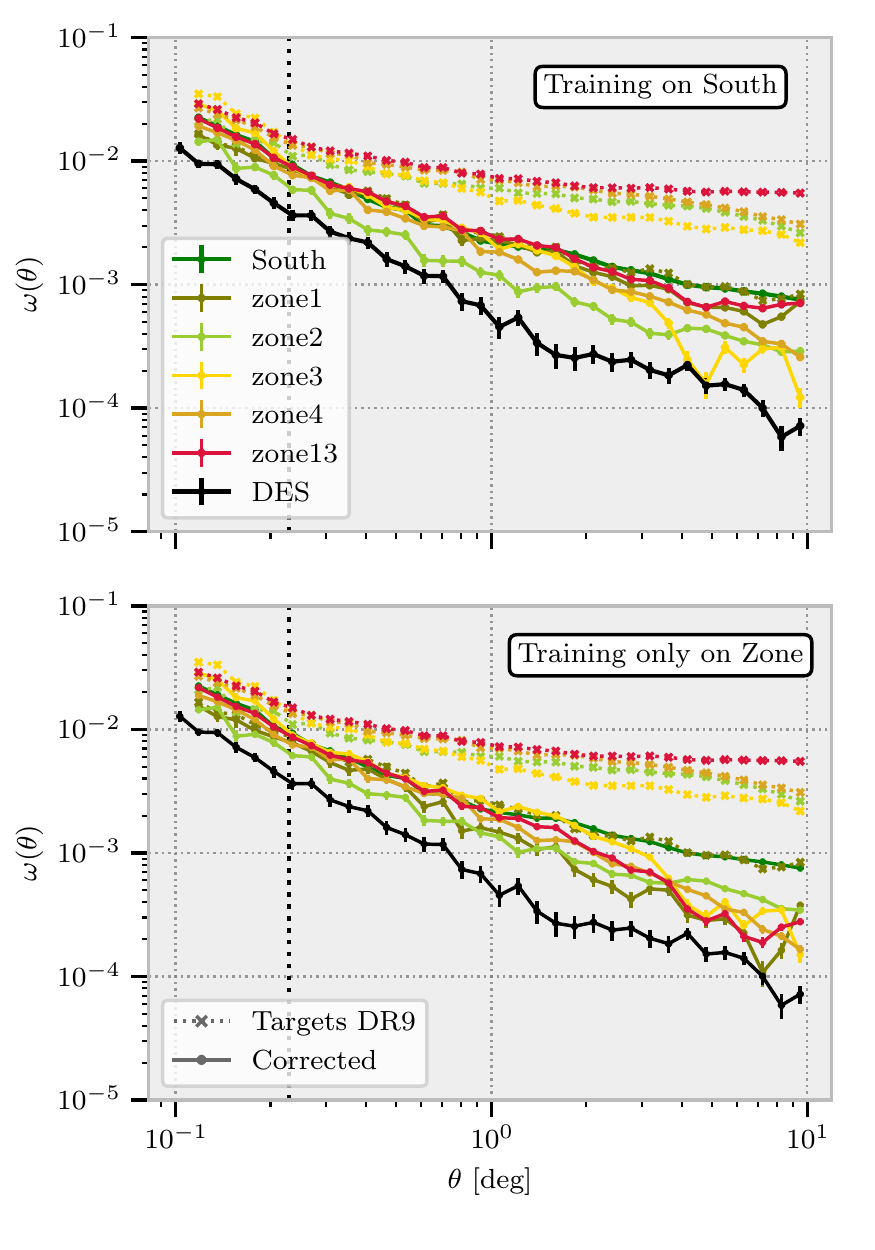}
    \centering
    \caption{The angular correlation for the four zones which are represented in Figure~\ref{fig:zone_repartition}.  Dotted lines are the correlations of DR9 targets,  the solid ones are for the corrected targets.  For comparison,  we also plot the correlation in the South region (green line) and in the DES region (black line) which is known as the least contaminated region.  On the top panel,  corrections are calculated on all the South footprint.  On the bottom panel,  corrections are calculated in each zone independently. }
    \label{fig:ang_corr_zone_south}
\end{figure}

The angular correlation functions for these different zones are shown in Figure~\ref{fig:ang_corr_zone_south}.  \rc{Here each mitigation is performed with the RF method.} The top panel shows the correction done with training on all the South footprint. The bottom panel shows the correction where the training was done individually on each consider zone.  We add also the angular correlation for the corrected targets in South (in green) and in DES (in black) for reference.  The correlations of the corrected targets on each zone are lower than on all the South footprint but it is not the average on zones due to the missing cross-terms.

\rc{The excess of correlation at small scales cannot be removed by our method. It is due to either stars or non-considered features. The part causes by a non-considered feature is left for a future study with the spectroscopic sample. The excess causes by the stellar contamination will be removed with the spectroscopic data as explained in Sec.~\ref{sec:restrictive}. } Hence,  we will discuss the excess of correlation at large scales compared to the correlation in DES.

The correction estimated with the training on all South footprint is not sufficient to recover the same level of correlation as in DES after mitigation for all zones.  The correction in zone 2 is more efficient than in the other zones since the Sagittarius Stream feature separates the Sagittarius Stream from the rest during the training. This zone does not recover the same angular correlation at large scales as in DES since the Sagittarius Stream feature is built as the spatial average of candidates stars suppressing the angular correlation information contained in this feature.  

The angular correlation in zone 1 without systematic mitigation is lower than those in the three other zones since this region does not show strong over-density.  However,  the systematic mitigation is inefficient in this region when the training is done on all the South footprint.  This is not the case when we apply the mitigation from the training only on zone 1.  Our method is unable to extract the correct information for this zone in all the South.  Zone 4 and zone 13 are better when the training is performed only on each zone,  \rc{recovering a level of correlation that is in more agreement with that of DES at large scales.}

\begin{figure}
    \includegraphics[scale=1]{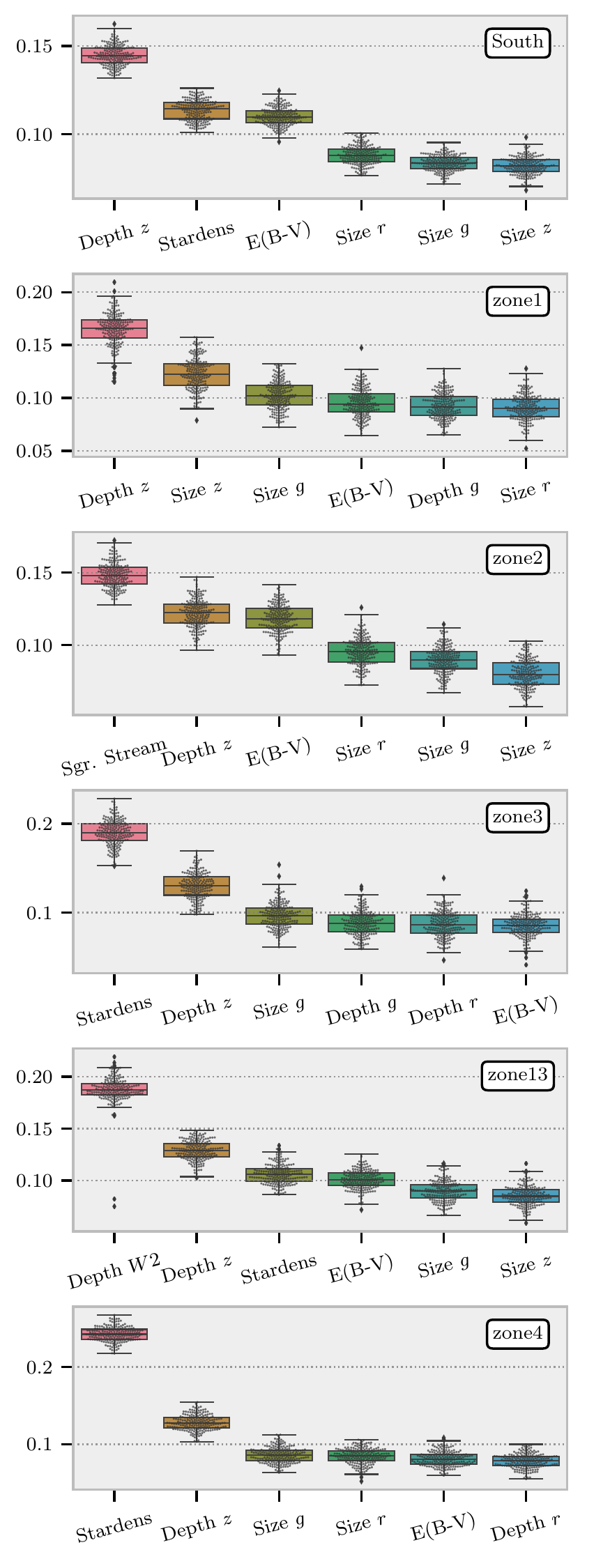}
    \centering
    \caption{Same as Figure~\ref{fig:feature_importance} but the feature importance plotted are estimated with the training only on the consider zone.  The different behaviors on each zone are highlighted by the different features which are qualified as important.  In particular,  the density of stars from the galactic plane prevents the regression to learn correctly the role of the PSF Depth $W2$ as expected. }
    \label{fig:feature_importance_zone}
\end{figure}

\rc{The inefficiency of the regression in all the South can be explained by the numerous stars in this region which bias the information about the observational systematic effects preventing the regression to learn correctly the true observational features dependency of the relative target density. } For instance,  the PSF Depth $W2$ is not characterized as an important feature (cf.  Sec.~\ref{sec:importance_feature}) while it explains the lower density observes around the anti-galactic pole.  Figure~\ref{fig:feature_importance_zone} shows the feature importance estimated with the training on the consider zone only.  The feature importance for all the zone used are plotted.  PSF Depth $W2$ \rc{is almost uniform (cf.  Figure~\ref{fig:feature_map} for its distribution) in zone 1 and zone 3 and so does not appear as an important feature when the training is done in each region. } However,  it is found as the most important feature in the training on zone 13 illustrating that the stars in zone 2 and in zone 4 bias the training, \eg the fluctuation of the relative density in the function of the PSF Depth $W2$ in zone 2 is masked by the presence of numerous stars from the Sagittarius Stream. 

\rc{Even if the training in an individual zone is more effective, especially at large scales than the training in all the South, we do not recover the same level correlation as in DES at intermediate scales. This can be explained still by the presence of numerous stars in each zone. The stellar contamination is mixed with the imaging systematics and the two effects are not easily separable. Then, the excess of correlation can be caused either by the stellar contaminant or by an additional unconsidered imaging feature.}

This analysis could be performed once the spectroscopic survey is done since the stellar contamination will vanish. \rc{The impact of imaging features can be then studied without any significant bias. If the set of features introduced during this analysis contains all the information, our systematic mitigation method will be able to correctly learn the true dependence on the observational features as in DES or the North and recover the correct angular correlation.}

\bsp	
\label{lastpage}
\end{document}